\documentclass[twocolumn,showpacs,preprintnumbers,amsmath,amssymb]{revtex4}
\usepackage{epsfig}
\usepackage{dcolumn}
\usepackage{bm}

\newcommand{\remove}[1]{}
\newcommand{\dd}{\mathrm{d}}

\def\be{\begin{equation}}
\def\ee{\end{equation}}

\newcommand{\beq}{\begin{equation}}
\newcommand{\eeq}{\end{equation}}
\newcommand{\beqa}{\begin{eqnarray}}
\newcommand{\eeqa}{\end{eqnarray}}

\renewcommand{\pl}{\partial}

\newcommand{\vu}{{\bf u}}
\newcommand{\vv}{{\bf v}}

\newcommand{\vx}{{\bf x}}

\renewcommand{\vr}{{\bf r}}

\newcommand{\vF}{{\bf F}}

\newcommand{\tg}{{\tilde{g}}}

\newcommand{\cG}{{\cal G}}
\newcommand{\cH}{{\cal H}}

\newcommand{\bea}{\begin{array}}
\newcommand{\ea}{\end{array}}

\newcommand{\MPl}{M_{\rm Pl}}

\begin{document}

\title{Small-scale Nonlinear Dynamics of K-mouflage Theories}

\author{Philippe Brax}
\affiliation{Institut de Physique Th\'eorique,\\
CEA, IPhT, F-91191 Gif-sur-Yvette, C\'edex, France\\
CNRS, URA 2306, F-91191 Gif-sur-Yvette, C\'edex, France}
\author{Patrick Valageas}
\affiliation{Institut de Physique Th\'eorique,\\
CEA, IPhT, F-91191 Gif-sur-Yvette, C\'edex, France\\
CNRS, URA 2306, F-91191 Gif-sur-Yvette, C\'edex, France}
\vspace{.2 cm}

\date{\today}
\vspace{.2 cm}

\begin{abstract}
We investigate the small-scale static configurations of K-mouflage models defined by a general function $K(\chi)$ of the kinetic terms.
The fifth force is screened by the nonlinear K-mouflage mechanism if $K'(\chi)$ grows sufficiently
fast for large negative $\chi$. In the general non-spherically symmetric case, the fifth force
is not aligned with the Newtonian force.
For spherically symmetric static matter density profiles,
we show that the results depend on the potential function $W_{-}(y) = y K'(-y^2/2)$, i.e. $W_{-}(y)$ must be monotonically increasing to $+\infty$ for $y \geq 0$
to guarantee the existence of a single solution throughout space for any matter density profile.
Small radial perturbations around these static profiles propagate as travelling waves with
a velocity greater than the speed of light. Starting from vanishing initial conditions for the scalar field and for a time-dependent matter density corresponding to the formation of an overdensity, we numerically check that the scalar field converges to the static solution.
If $W_{-}$ is bounded, for high-density objects there are no static solutions throughout space,
but one can still define a static solution restricted to large radii. Our dynamical study shows that
the scalar field relaxes to this static solution at large radii, whereas
spatial gradients keep growing with time at smaller radii.
If $W_{-}$ is not bounded but non-monotonic, there is an infinite number of discontinuous
static solutions. However, the Klein-Gordon equation is no longer a well-defined
hyperbolic equation, which leads to complex characteristic speeds and exponential instabilities.
Therefore, these discontinuous static solutions are not physical and these models are
not theoretically sound.
Such K-mouflage scenarios provide an example of theories that can appear viable at the
cosmological level, for the cosmological background and perturbative analysis,
while being meaningless at a nonlinear level for small-scale configurations.
This shows the importance of small-scale nonlinear analysis of screening models.
All healthy K-mouflage models should satisfy $K'>0$ and $W_{\pm}(y) = y K'(\pm y^2/2)$
are monotonically increasing to $+\infty$ when $y \geq 0$.

\keywords{Cosmology \and large scale structure of the Universe}
\end{abstract}

\pacs{98.80.-k} \vskip2pc

\maketitle

\section{Introduction}
\label{sec:Introduction}

Theories with second-order equations of motions involving a single scalar field and a coupling to matter can be subject to three different screening mechanisms whereby the scalar interaction is screened in dense environments\cite{Khoury:2013tda}.
The Vainshtein mechanism \cite{Vainshtein:1972sx}, which is present in DGP models \cite{Dvali2000} and Galileons \cite{Nicolis:2008in}, was the first to be uncovered and involves models where higher-order derivatives appear in the action. The scalar field is screened in regions of space where the local curvature is large enough. The chameleon mechanism \cite{Khoury:2003aq,Khoury:2003rn} involves another type of nonlinearities and plays a role in models where a nonlinear potential can for instance generate the late-time acceleration of the  expansion of the Universe \cite{Brax:2004qh}. Screening takes places where the scalar itself is small compared to the ambient Newtonian potential. Finally, the K-mouflage mechanism \cite{Babichev:2009ee,Brax:2012jr} is present in models where the Lagrangian involves an arbitrary function $K(\chi)$ of the kinetic terms. Screening appears in regions of space where the gravitational acceleration is large enough. The three types of screening have different cosmological properties and lead to different behaviors on the very large scale structures of the Universe. For Galileon models, the background cosmology is defined by a late-time attractor \cite{Li:2013tda}. Hence, the cosmological configurations converge rapidly to the $\Lambda$-CDM paradigm in the recent past of the Universe. Moreover, large clusters of galaxies are screened leading to small deviations of the growth of structure from $\Lambda$-CDM, even on the largest scales. On the other hand, for chameleon models such as $f(R)$ \cite{Hu:2007nk}, the background follows $\Lambda$-CDM \cite{Brax:2012gr}, and the growth of structure is only sensitive to modified gravity on intermediate quasi-linear scales. Finally, for K-mouflage, the background cosmology \cite{Brax:2014aa} shows a host of different behaviors. In particular, for ghost-less models with a polynomial kinetic function, the effective equation of state shows a singularity in the recent past and crosses the Phantom divide. This has no strange effect on the dynamics though, as the Hubble rate squared remains positive definite. Moreover, large galaxy clusters are not screened and can show deviations from $\Lambda$-CDM in the halo mass function for instance \cite{Brax:2014aa}.

Here we investigate the small-scale static properties of K-mouflage theories. We find that the resulting field configurations depend on a potential $W_{-}(y)=y K'(-y^2/2)$. When this potential is
monotonically increasing to $+\infty$ over $y \geq 0$, the scalar modification of gravity is attractive,
and we find that a well-defined scalar-field static profile exists for any matter density profile.
Small perturbations around these configurations are travelling waves with a speed greater than the speed of light. We also show that the static profile emerges dynamically as a long-time solution of the dynamical evolution of the scalar field from vanishing initial conditions. This is the case for the cubic interaction model in the kinetic terms, e.g. $K(\chi) = -1 + \chi + \chi^3$, with a bounded from below Lagrangian, or the wrong-sign DBI$^{+}$ models presented in \cite{Brax:2012jr,Burrage2014}. On the other hand, when the potential is bounded or has several extrema, the dynamics are pathological. For bounded potentials, the convergence to a static solution only happens when the object is not screened. When it is screened, i.e. when its size is smaller than its K-mouflage radius, the convergence to a static solution can only be reached at large radii, outside of the K-mouflage radius, while scalar gradients keep growing with time inside the K-mouflage radius. This case corresponds to the DBI$^{-}$ models, which therefore must be altered within the K-mouflage radii of screened objects in order to make sense.  When the potential has several extrema, the Klein-Gordon equation is a not a well-defined hyperbolic equation. Instabilities associated to complex characteristic speeds occur and the evolution with time is no longer well defined (as one
encounters an elliptic problem that requires boundary conditions at late times, instead
of a Cauchy problem).

In Sec.II, we recall the definition and main properties of K-mouflage models.
In Sec.III, we describe the nonlinear K-mouflage screening mechanism and spherically
symmetric static configurations, as well as the dynamics of radial perturbations.
In Sec.IV, we consider generalized static solutions that can still be defined when
$W_{-}(y)$ is not monotonically increasing to $+\infty$.
In Sec.V, we analyze the dynamics of these configurations and their generation from vanishing initial conditions.
In Sec.VI, we summarize the cosmological and static properties of K-mouflage theories. We then conclude.

\section{K-mouflage}
\label{sec:K-mouflage}

\subsection{Definition of the model}
\label{sec:definition-model}

As in \cite{Brax:2014aa,Brax:2014ab}, we consider scalar field models where the action in the
Einstein frame has the form
\beqa
S & = & \int \dd^4 x \; \sqrt{-g} \left[ \frac{\MPl^2}{2} R + {\cal L}_{\varphi}(\varphi)
\right] \nonumber \\
&& + \int \dd^4 x \; \sqrt{-\tg} \, {\cal L}_{\rm m}(\psi^{(i)}_{\rm m},\tg_{\mu\nu}) ,
\label{S-def}
\eeqa
where $\MPl=1/\sqrt{8\pi\cG}$ is the reduced Planck mass,
 $g$ is the determinant of the metric tensor $g_{\mu\nu}$, and
$\psi^{(i)}_{\rm m}$ are various matter fields.
The additional scalar field $\varphi$ is explicitly coupled to matter through the
Jordan-frame metric $\tg_{\mu\nu}$, which is given by the conformal rescaling
\beq
\tg_{\mu\nu} = A^2(\varphi) \, g_{\mu\nu} ,
\label{g-Jordan-def}
\eeq
and $\tg$ is its determinant.
We have already considered various canonical scalar field models in previous
works \cite{BraxPV2012,BraxPV2013}, with
${\cal L}_{\varphi} = - (\pl\varphi)^2/2 - V(\varphi)$.
In this paper, we consider models with a non-standard kinetic term
\beq
{\cal L}_{\varphi}(\varphi) = {\cal M}^4 \, K(\chi) \;\;\; \mbox{with} \;\;\;
\chi = - \frac{1}{2{\cal M}^4} \pl^{\mu}\varphi\pl_{\mu}\varphi .
\label{K-def}
\eeq
[We use the signature $(-,+,+,+)$ for the metric.]
To focus on the behavior associated with the non-standard kinetic term $K$, we
do not add a potential $V(\varphi)$ or a mixed dependence $K(\varphi,\chi)$ on the
the field value and the derivative terms.
Here, ${\cal M}^4$ is an energy scale that is of the order of the current energy density,
(i.e., set by the cosmological constant), to recover the late-time accelerated expansion
of the Universe.
Thus, the canonical cosmological behavior, with a cosmological constant
$\rho_{\Lambda} = {\cal M}^4$, is recovered at late time in the weak-$\chi$ limit if we have:
\beq
\chi \rightarrow 0 : \;\;\; K(\chi) \simeq -1 + \chi + ... ,
\label{K-chi=0}
\eeq
where the dots stand for higher-order terms.
On the other hand, the property
\beq
K'(0) =1
\label{Kp0-1}
\eeq
defines the normalization of the scalar field $\varphi$ and can be chosen without loss of generality,
assuming $K'(0)$ is positive and finite.
This sign ensures that at low $\chi$ we recover a standard kinetic term with the correct sign
and we avoid ghosts.

In the cosmological regime, associated with (large) positive $\chi$, we must have $K'(\chi)>0$
to avoid ghost instabilities \cite{Brax:2014aa}.
Otherwise, quantum vacuum instabilities produce too much radiation as compared with observational
bounds on the gamma-ray spectrum and we must introduce a UV cutoff at a very low energy scale
[a few keV or eV, depending on the form of the kinetic function $K(\chi)$], which makes the model
very contrived.
In cases where $K'(\chi)>0$ over the whole positive real axis, $\chi \geq 0$, the cosmological
background evolves from very large positive values of $\bar\chi$ at early times to $\bar\chi\rightarrow 0$
at late times, $t\rightarrow \infty$.
In cases where $K'(\chi_*)=0$ for some positive value $\chi_*>0$, and $K'(\chi)>0$ for $\chi>\chi_*$,
the cosmological background $\bar\chi$ rolls down from $+\infty$ to $\chi_*$ as time increases,
and there are no ghosts in the perturbative regime around the background \cite{Brax:2014aa}.

However, models where $K'(\chi)$ has a zero $\chi_*$ on the real axis, whether it is negative or positive,
are badly behaved. Indeed, while the cosmological regime probes the domain $\chi > \max(0,\chi_*)$,
where $K'>0$, the small-scale quasistatic regime probes the domain $\chi<0$, and highly nonlinear
screened objects correspond to $\chi \ll -1$.
Then, as the scalar field evolves from the large-scale cosmological regime to the small-scale
screened regime, it goes through the point $\chi_*$. At this point, the fifth force diverges, as seen from
the expression (\ref{Geff-1}) of the effective Newtonian constant obtained below, or from the
fifth-force expression (\ref{F-fifth-force}).
This also gives infinite propagation speeds $c_s^2$ for the scalar field, with a change of sign of $c_s^2$
that signals divergent instability growth rates, as seen from the expression (\ref{c-speed-1}) obtained
below.
To avoid these divergences and instabilities, we require that $K'(\chi)$ does not change sign.
Therefore, in this paper we focus on models where $K'(\chi)>0$ for all values of $\chi$.

\subsection{Specific models}
\label{sec:specific-models}

In \cite{Brax:2014aa,Brax:2014ab}, we considered for the coupling function $A(\varphi)$ the
simple power laws,
\beq
n \in {\mathbb N} , \;\; n \geq 1 : \;\;\;
A(\varphi) = \left( 1 + \frac{\beta\varphi}{n \MPl} \right)^n ,
\label{A-power-1}
\eeq
and the exponential limit for $n \rightarrow +\infty$,
\beq
A(\varphi) = e^{\beta \varphi/\MPl} .
\label{A-exp-1}
\eeq
Without loss of generality, we normalize the field $\varphi$ by the appropriate
additive constant so that $A(0)=1$.
In fact, in the regime that we consider in this paper, we only need the first-order
expansion $A(\varphi) \simeq 1 + \beta\varphi/M_{\rm Pl}$, so that all coupling functions
(\ref{A-power-1}) and (\ref{A-exp-1}) coincide for our purposes.
The action (\ref{S-def}) is invariant with respect to the transformation
$(\varphi,\beta) \rightarrow (-\varphi,-\beta)$, therefore we can choose $\beta >0$.

For the kinetic function $K(\chi)$, we considered in \cite{Brax:2014aa,Brax:2014ab}
the polynomials
\beq
m \in {\mathbb N} , \;\; m \geq 2 : \;\;\;
K(\chi) = -1 + \chi + K_0 \, \chi^m ,
\label{K-power-1}
\eeq
and we focused on the low-order cases $m=2$ and $3$.
In this paper, we focus on the case $K_0=1, m=3$, for numerical applications.

In addition, we consider models of the Dirac-Born-Infeld (``DBI'') type, with a non-standard
sign as in \cite{Burrage2014},
\beq
\mbox{DBI}^{+} \; : \;\;\; K(\chi)= \sqrt{1+2\chi} - 2  ,
\label{K-DBI+def}
\eeq
and with the standard sign,
\beq
\mbox{DBI}^{-} \; : \;\;\; K(\chi)= -\sqrt{1-2\chi} .
\label{K-DBI-def}
\eeq

\subsection{Equation of motion of the scalar field}
\label{sec:equation-motion-scalar}

The Klein-Gordon equation that governs the dynamics of the scalar field
$\varphi$ is obtained from the variation of the action (\ref{S-def}) with respect to
$\varphi$. This gives \cite{Brax:2014aa,Brax:2014ab}
\beq
\frac{1}{\sqrt{-g}} \pl_{\mu} \left[ \sqrt{-g} \; g^{\mu\nu} \pl_{\nu} \varphi \; K' \right] =
\frac{\dd\ln A}{\dd\varphi} \; \rho_E  ,
\label{KG-1}
\eeq
where $\rho_E=- g^{\mu\nu}T_{\mu\nu}$ is the Einstein-frame matter density,
and we note with a prime $K'=\dd K/\dd\chi$.
In the limit where the metric fluctuations $\Psi$ are small, $\Psi \ll c^2$, which applies to
cosmological and galactic scales, this reads as
\beq
\frac{1}{a^3} \frac{\pl}{\pl t} \left( a^3 \frac{\pl \varphi}{\pl t} \; K' \right) - \frac{1}{a^2} \nabla \cdot (\nabla\varphi \; K' ) = - \frac{\dd A}{\dd \varphi} \, \rho ,
\label{KG-2}
\eeq
with $\chi = 1/(2{\cal M}^4) \left[ (\pl\varphi/\pl t)^2 - 1/a^2 (\nabla\varphi)^2 \right]$,
where we use comoving coordinates and $\rho=A^{-1} \rho_E$ is the conserved matter density
\cite{Brax:2014aa,Brax:2014ab}.
The cosmological evolution associated with K-mouflage models was studied in previous papers
of this series, for both the background dynamics \cite{Brax:2014aa}
and the formation of large-scale structures \cite{Brax:2014ab}.
In this paper, we are interested in the small-scale nonlinear regime where screening mechanisms
come into play and lead to a convergence back to General Relativity for galactic and astrophysical
systems (e.g., the Solar System).
Then, going to physical space coordinates, $\vr = a \vx$, considering time scales that are much
smaller than the cosmological time scale [i.e., $a(t)$ is almost constant and $H=\dot{a}/a\approx 0$] and densities
that are much greater than the mean universe density, the Klein-Gordon equation
(\ref{KG-2}) becomes
\beq
\frac{\pl}{\pl t} \left( \frac{\pl \varphi}{\pl t} \; K' \right)
- c^2 \nabla_{\vr} \cdot (\nabla_{\vr} \varphi \; K' ) = - \frac{\beta\rho}{M_{\rm Pl}} ,
\label{KG-3}
\eeq
with
\beq
\chi = \frac{1}{2{\cal M}^4} \left[ \left( \frac{\pl \varphi}{\pl t} \right)^2
 - c^2 (\nabla_{\vr}\varphi)^2 \right] ,
\label{chi-r}
\eeq
where we have explicitly written the factors of $c^2$.

On the right-hand-side of Eq.(\ref{KG-3}) we have used the approximation
$A(\varphi) \simeq 1+\beta\varphi/M_{\rm Pl}$, which holds for the power-law and
exponential models (\ref{A-power-1})-(\ref{A-exp-1}) as long as $\beta | \varphi | /M_{\rm Pl} \ll 1$.
From Eq.(\ref{KG-3}), this corresponds to the regime
\beq
\left| \frac{\beta \varphi}{M_{\rm Pl}} \right| \sim
\frac{\beta^2}{K'} \; \frac{\rho}{\bar\rho} \; \frac{r^2}{r_H^2} \ll 1 ,
\label{phi-small-1}
\eeq
where $\bar\rho$ is the mean universe matter density and $r_H = c t_H$ is the Hubble radius.
The K-mouflage models that we consider typically have $\beta \lesssim 1$ and $K' \gtrsim 1$ on the
positive branch, $\chi>0$, to satisfy cosmological constraints \cite{Brax:2014aa,Brax:2014ab}.
For generic cases, $K'$ remains of order unity or greater on the negative branch, $\chi_0<0$,
 where we focus on small-scale static configurations.
Thus, the ratio $\beta^2/K'$ is typically smaller than unity.
For a typical cluster of galaxies, we have $\rho/\bar\rho \sim 200$ and $r/r_H \sim 1/3000$,
which gives $\rho r^2/{\bar\rho} r_H^2 \sim 10^{-5}$.
For the Solar System, up to the Jupiter orbit, we have $\rho/\bar{\rho} \sim 10^{20}$ and
$r/r_H \sim 10^{-14}$, which gives $\rho r^2/{\bar\rho} r_H^2 \sim 10^{-8}$.
For the Sun, we have $\rho/\bar{\rho} \sim 4 \times 10^{29}$ and
$r/r_H \sim 5 \times 10^{-18}$, which gives $\rho r^2/{\bar\rho} r_H^2 \sim 10^{-5}$,
while for the Earth we obtain $\rho r^2/{\bar\rho} r_H^2 \sim 10^{-9}$.
Therefore, the condition (\ref{phi-small-1}) is satisfied in all cases of interest considered in
this paper, from planets to galaxies and clusters, and we can truncate the function $A(\varphi)$ to
the first-order term $A(\varphi) \simeq 1 + \beta\varphi/M_{\rm Pl}$.

\subsection{Poisson equation}
\label{sec:Poisson}

The Einstein-frame metric potential $\Psi_{\rm N}$ is given by the modified Poisson equation
\cite{Brax:2014ab}
\beq
\frac{1}{a^2} \nabla^2 \Psi_{\rm N} = 4\pi\cG (\delta\rho_E + \delta \rho_{\varphi} ) ,
\label{Poisson1}
\eeq
where $\rho_{\varphi} = - {\cal M}^4 K + (\pl\varphi/\pl t)^2 K'$ is the scalar field energy density.
In this paper we focus on small nonlinear scales, associated with galactic or astrophysical
objects.
In this regime, using the property (\ref{phi-small-1}),
hence $| A-1 | \simeq \beta | \varphi | /M_{\rm Pl} \ll 1$,
we can write $\delta\rho_E \simeq \delta \rho \simeq \rho$, as we consider high-density objects
with $\rho \gg \bar\rho$.
At low redshifts, we also have $\bar\rho_{\varphi} \simeq {\cal M}^4 \sim \bar\rho$.
If $\chi$ and $K$ remain of order unity, we have
$\rho_{\varphi} \sim \bar\rho_{\varphi} \sim \bar\rho$.
If $\chi \ll -1$, i.e. we are in the nonlinear static regime of the Klein-Gordon equation
(\ref{KG-3}), we obtain:
\beq
\chi \sim \frac{\beta^2 \rho^2 r^2}{K'^2 {\bar\rho}^2 r_H^2} \;\;\;
\mbox{and} \;\;\; \frac{\rho_{\varphi}}{\rho} \sim \frac{\bar\rho K' \chi}{\rho} \sim
\frac{\beta^2}{K'} \; \frac{\rho}{\bar\rho} \; \frac{r^2}{r_H^2} \ll 1 ,
\eeq
where we used Eq.(\ref{phi-small-1}).
Therefore, in the small-scale high-density regime the Poisson equation simplifies
and we recover the standard form
\beq
\nabla_{\vr}^2 \Psi_{\rm N} = 4\pi\cG \rho .
\label{Poisson2}
\eeq

\subsection{Euler equation}
\label{sec:Euler}

The pressureless Euler equation which describes the dark matter flow on large cosmological scales
reads as \cite{Brax:2014ab}
\beq
\frac{\pl \vv}{\pl\tau} + (\vv\cdot\nabla) \vv + \left( \cH + \frac{\pl \ln A}{\pl \tau} \right) \vv =
- \nabla (  \Psi_{\rm N} + \ln A ) ,
\label{Euler-1}
\eeq
where $\tau= \int \dd t/a$ is the conformal time, $\cH=\dot{a}=\dd\ln a/\dd\tau$ the conformal
expansion rate, and $\vv=a \dot{\vx}$ the peculiar velocity.
Going to physical coordinates in the small-scale regime, and considering time scales that are much
smaller than the cosmological time scale, this yields
\beq
\frac{\pl \vu}{\pl t} + (\vu\cdot\nabla_{\vr}) \vu =
- \nabla_{\vr} \left( \Psi_{\rm N} + \frac{c^2\beta\varphi}{M_{\rm Pl}} \right) ,
\label{Euler2}
\eeq
where we have explicitly written  the factors of $c^2$ and we have used the property (\ref{phi-small-1}).
Here $\vu=\vv+H\vr=\dot{\vr}$ is the physical velocity.

\section{The K-mouflage  mechanism}
\label{sec:screening}

\subsection{Static case}
\label{sec:static}

In the static case, the Klein-Gordon equation (\ref{KG-3}) becomes
\beq
\nabla_{\vr} \cdot (\nabla_{\vr} \varphi \; K' ) =  \frac{\beta\rho}{c^2 M_{\rm Pl}} ,
\label{KG-static-1}
\eeq
with $\chi= -c^2 (\nabla_{\vr}\varphi)^2/(2{\cal M}^4)$,
and the comparison with the Poisson equation (\ref{Poisson2}) gives the relation
\beq
\nabla_{\vr} \varphi \; K' = \frac{2\beta M_{\rm Pl}}{c^2} \left( \nabla_{\vr} \Psi_{\rm N}
+ \nabla_{\vr} \times {\vec\omega} \right) ,
\label{KG-omega-curl}
\eeq
where ${\vec\omega}$ is a divergence-less potential vector
(which must be determined along with $\varphi$).
In general configurations, $\nabla_{\vr} \varphi \; K'$ is not curl-free (because of the
spatial dependence of $K'$) and ${\vec\omega}$ is nonzero.
However, in special configurations it can be shown to vanish.
In particular, this is automatically the case in spherically symmetric systems,
which we study in Sec.~\ref{sec:static-spherical} below,
as $\nabla_{\vr}\times [ \omega(r) {\bf e}_{\vr} ] = 0$.
In the general case, we obtain from Eq.(\ref{KG-omega-curl}) the
relation
\beq
\nabla_{\vr} \times \left( \frac{\nabla_{\vr} \times {\vec\omega}}{K'} \right) =
- \nabla_{\vr} \left( \frac{1}{K'} \right) \times \nabla_{\vr} \Psi_{\rm N} .
\label{omega-Kp-Psi}
\eeq

Let us briefly consider the case of a localized nonlinear fluctuations.
On large scales, far from the central nonlinearities, the gravitational force
$\nabla_{\vr}\Psi_{\rm N}$ vanishes as $1/r^2$.
Then, for models that have the low-$\chi$ expansion (\ref{K-chi=0}),
we are in the linear regime with:
\beq
\mbox{weak field:} \;\;\; \varphi = \frac{2\beta M_{\rm Pl}}{c^2} \, \Psi_{\rm N} ,
\;\;\; {\vec\omega} =0 ,
\label{weak}
\eeq
using the boundary conditions $\varphi \rightarrow 0$ and $\Psi_{\rm N} \rightarrow 0$
at infinity. Next, we can solve Eq.(\ref{KG-omega-curl}) for $\nabla_{\vr}\varphi$
as a perturbative expansion over $\Psi_{\rm N}$.
Because $\chi \propto (\nabla_{\vr}\varphi)^2$, only odd-order terms are nonzero.
The first-order term is given by Eq.(\ref{weak}), and at third order we obtain that
$\nabla_{\vr}\varphi^{(3)}$ and $\nabla_{\vr}\times {\vec\omega}^{(3)}$ are the potential
and rotational parts of $\nabla_{\vr}\Psi_{\rm N} (\nabla_{\vr}\Psi_{\rm N})^2$.

The fifth force, which can be read from Eq.(\ref{Euler2}), is
\beq
\vF_{\varphi} \equiv - \frac{\beta c^2}{M_{\rm Pl}} \nabla_{\vr}\varphi =
- \frac{2\beta^2}{K'} \left( \nabla_{\vr} \Psi_{\rm N} + \nabla_{\vr} \times {\vec\omega} \right) .
\label{F-fifth-force}
\eeq
The K-mouflage screening mechanism relies on the fact that in the nonlinear regime
the factor $K'$ can be large, which suppresses the fifth force as compared with the
Newtonian gravity, $\vF_{\rm N} = -\nabla_{\vr}\Psi_{\rm N}$, with
$|  \vF_{\varphi} | \sim | \vF_{\rm N} / K' |$.
This also applies to the rotational part, since from Eq.(\ref{omega-Kp-Psi}) we have
the scaling $| \nabla_{\vr} \times \vec\omega | \sim | \nabla_{\vr} \Psi_{\rm N} |$.

In the general case (i.e., when the density field is not spherically symmetric),
Eqs.(\ref{KG-omega-curl}) and (\ref{F-fifth-force}) imply that the gradient of the scalar field,
$\nabla_{\vr}\varphi$, and the fifth force, $\vF_{\varphi} \propto \nabla_{\vr}\varphi$,
are not aligned with the Newtonian force $\nabla_{\vr}\Psi_{\rm N}$.
This is because the relationship (\ref{KG-omega-curl}) between
$\nabla_{\vr}\varphi$ and $\nabla_{\vr}\Psi_{\rm N}$ involves an additional divergence-less
field ${\vec\omega}$ that arises from the rotational part of $\nabla_{\vr}\Psi_{\rm N}/K'$.
This makes the study of several-body problems complicated, with non-parallel Newtonian
force and fifth force and an additional component $\nabla_{\vr} \times {\vec\omega}$
in Eq.(\ref{F-fifth-force}).

\subsection{Static spherical profile}
\label{sec:static-spherical}

We are interested in the dynamics of test objects in the background of a denser and compact body. This body could be a star, a galaxy or a gas cloud. For simplicity, we restrict ourselves to spherical configurations, so that the solenoidal term
$\nabla_{\vr} \times {\vec\omega}$ in Eq.(\ref{KG-omega-curl}) vanishes.
We consider a spherical matter distribution $\rho(r)$ with the mass profile
\be
M(r)= \int_0^r \dd r' \; 4\pi r'^2 \, \rho (r')  .
\ee
The Klein-Gordon equation (\ref{KG-static-1}) gives, using Stokes theorem,
\be
\frac{\dd\varphi}{\dd r} \; K' \left(- \frac{c^2}{2{\cal M}^4} \left(\frac{\dd\varphi}{\dd r}\right)^2 \right) = \frac{\beta M(r)}{c^2 M_{\rm Pl} 4\pi r^2} .
\label{KG-4}
\ee
As in \cite{Brax:2014aa,Brax:2014ab}, we define a ``K-mouflage screening radius'' $R_K$ by
\beq
R_K = \left( \frac{\beta M}{4\pi c M_{\rm Pl} {\cal M}^2} \right)^{1/2} ,
\label{RK-def}
\eeq
where $M=M(R)$ is the total mass of the object of radius $R$.
Then, introducing the rescaled dimensionless variables $x=r/R_K$, $m(x)= M(r)/M$, $\phi(x)= \varphi(r)/\varphi_K$, with
\beq
\varphi_K = {\cal M}^2 R_K/c ,
\label{phiK-def}
\eeq
the integrated Klein-Gordon equation (\ref{KG-4}) reads as
\beq
\frac{\dd\phi}{\dd x} \, K'\left[ - \frac{1}{2} \left( \frac{\dd\phi}{\dd x} \right)^2 \right] = \frac{m(x)}{x^2} .
\label{KG-5}
\eeq
Then, it is convenient to define the potential function $W_{-}(y)$, which will play a crucial role in the following, by
\be
W_{-}(y)= y K'(-y^2/2) ,
\label{Wy-def}
\ee
(where the subscript ``-'' recalls the minus sign in the argument of $K'$), so that Eq.(\ref{KG-5}) becomes
\be
W_{-}(y)=\frac{m(x)}{x^2}  \;\;\; \mbox{with} \;\;\; y(x) = \frac{\dd\phi}{\dd x} .
\label{KG-W-1}
\ee

If the kinetic function $K(\chi)$ obeys the weak-$\chi$ behavior (\ref{K-chi=0}), we have
\beq
y \rightarrow 0 : \;\;\; W_{-}(y) \simeq y + ... ,
\label{W-small-y}
\eeq
where the dots stand for higher-order odd terms.
The standard kinetic term corresponds to $W_{-}(y)=y$ without nonlinear corrections, which
describe the distinctive K-mouflage nonlinear features.

For a given matter profile $m(x)$, Eq.(\ref{KG-W-1}) provides the radial profile of the scalar field
$\phi(x)$. More precisely, assuming this equation can be inverted, it yields the first derivative
$\dd\phi/\dd x = W_{-}^{-1}(m/x^2)$ at all positions $x$, hence $\phi(x)$ using the boundary
condition $\phi(\infty) = 0$ at infinity, in the vacuum far from the finite-size object.
In particular, using the low-$y$ expansion (\ref{W-small-y}), we obtain at large distances
[by definition $m(x)=1$ for $x>R/R_K$]:
\beq
x \gg 1 \;\; \mbox{and} \;\; x > \frac{R}{R_K} : \;\; \frac{\dd\phi}{\dd x} = y \simeq \frac{1}{x^2} ,
\;\; \phi(x) \simeq - \frac{1}{x} .
\label{phi-large-x}
\eeq
In the standard case, where $W_{-}(y)=y$, this solution is exact down to $x=R/R_K$.
Since we consider an object with a finite central matter density, we have $m(x) \propto x^3$
for $x \rightarrow 0$, and we obtain the small-radius behavior:
\beq
x \rightarrow 0 : \;\; \frac{\dd\phi}{\dd x} = y \propto x ,
\;\; \phi(x) \simeq \phi_0 + \phi_2 \frac{x^2}{2} + ... ,
\label{phi-small-x}
\eeq
where $\phi_0$ is the value of the scalar field at the center.

Thus, $W_{-}$ and $y=\dd\phi/\dd x$ vanish at both $x=0$ and $x \rightarrow \infty$.
On intermediate scales, $W_{-}$ is strictly positive. If the potential $W_{-}(y)$ is monotonically
increasing up to $\infty$ [as in the standard case where $W_{-}(y)=y$] the solution of
Eq.(\ref{KG-W-1}) is unique and well defined.

Regular examples would be for instance the power-law models
(\ref{K-power-1}) with:
\beq
{m \;\; \mbox{even and} \;\; K_0 < 0 , \;\;\; \mbox{or} \;\; m \;\; \mbox{odd and} \;\;  K_0 > 0 ,}
\label{K0-no-sing}
\eeq
which give
\beq
W_{-}(y) = y \left[ 1 + K_0 (-1)^{m-1} m (y^2/2)^{m-1} \right]  .
\label{W-K0-m}
\eeq
This function $W_{-}(y)$ is defined over the full real axis, it is monotonically increasing,
and it goes to $+\infty$ for $y \rightarrow +\infty$.
Then, the scalar field derivative $\dd\phi/\dd x=y$ can take arbitrarily large values.

Alternative models of the same class, with a monotonically increasing
$W_{-}(y)$ up to $+\infty$ over $y \geq 0$, correspond to cases where $W_{-}(y)$ diverges
at a finite value $y_->0$.
An example is provided by the DBI-like model (\ref{K-DBI+def})
studied in \cite{Burrage2014},
with a nonstandard sign. It corresponds to $W_{-}(y) = y/\sqrt{1-y^2}$,
which is monotonically increasing up to $+\infty$ over $0 \leq y < 1$.
Then, the scalar field derivative $\dd\phi/\dd x$ cannot be greater than $1$
(which also shows at once that the fifth force becomes negligible as compared
with Newtonian gravity in strong gradient regimes).

When the function $W_{-}(y)$ is not monotonically increasing up to $+\infty$ over some range
$[0,y_-[$ with $y_->0$ (including the case $y_-=+\infty$), there is no unique well-defined
continuous solution.
In particular, for sufficiently dense objects, if $W_{-}(y)$ is bounded there is no solution
that applies at all radii,
whereas if $W_{-}(y)$ is unbounded but not monotonic there are infinitely many discontinuous
solutions. We postpone the analysis of these cases to Sec.~\ref{sec:not-inverted} below.

Therefore, the condition for a well-defined and unique scalar field profile for any matter
overdensity is
\beq
\mbox{static solution:} \;\;\; W_{-}'(y)  \geq 0 \;\;\; \mbox{and} \;\;\; W_{-} \rightarrow +\infty ,
\label{condition-static-W}
\eeq
over a range $0 \leq y < y_-$, where $y_-$ can be finite or $+\infty$.
The condition $W_{-}'>0$ also reads in terms of the kinetic function $K$ as
\beq
\mbox{static solution:} \;\;\; K' > 0, \;\;\; K' + 2 \chi K'' \geq 0 ,
\label{condition-static-K}
\eeq
over the range $\chi_- < \chi \leq 0$, where $\chi_-=-y_-^2/2$ is either finite or $-\infty$.
The property $K'>0$ comes from the fact that starting at $K'=1$ at $x\rightarrow\infty$ and
$\chi\rightarrow 0$, $K'$ cannot change sign nor vanish as we move closer to the object
while $W_{-}$ increases as $m(x)/x^2$ because of the definition (\ref{Wy-def})
(i.e., $W_{-}$ cannot go through zero as it is always strictly positive).

\subsection{Corrections to Newton's law}
\label{sec:Newton-law}

We now focus on the simple case where Eq.(\ref{KG-W-1}) has a unique well-defined solution,
which corresponds to the models (\ref{K0-no-sing}), or more generally to models where $W_{-}(y)$
is monotonically increasing up to $+\infty$ over an interval $[0,y_-[$.
We discuss the corrections to Newton's law to check how the nonlinear K-mouflage screening
mechanism provides a convergence back to GR (or Newtonian gravity) on small astrophysical
scales.
A test particle outside the dense body evolves according to the non-relativistic equation
[see Eq.(\ref{Euler2})]
\be
\frac{\dd^2 \vr}{\dd t^2}= - \nabla_{\vr} \Psi_{\rm N}
- \frac{\beta c^2}{M_{\rm Pl}} \nabla_{\vr} \varphi .
\label{5th-force}
\ee
For a spherical body we can consider radial trajectories and the scalar field gradient is
given by Eq.(\ref{KG-5}), which can also be written as
\beq
\frac{\dd\varphi}{\dd r} = \frac{\varphi_K}{R_K} \; \frac{m(x)}{x^2 K'} .
\eeq
Outside the spherical body we have $m(x)=1$ and we obtain
in agreement with Eq.(\ref{F-fifth-force})
\beq
\frac{\beta c^2}{M_{\rm Pl}} \frac{\dd\varphi}{\dd r} = \frac{2 \beta^2 \cG M}{K' r^2}
= \frac{2\beta^2}{K'} \frac{\dd\Psi_{\rm N}}{\dd r} ,
\label{force-1}
\eeq
which gives the equation of motion
\beq
\frac{\dd^2 r}{\dd t^2} = - \frac{\cG M}{r^2} \; \left( 1+ \frac{2\beta^2}{K'} \right) .
\label{radial-1}
\eeq
This corresponds to an effective Newtonian constant
\beq
\cG^{\rm eff}(r) = \left( 1+ \frac{2\beta^2}{K'(\chi(r))} \right) \; \cG
\label{Geff-1}
\eeq
that depends on the distance from the central object.

From the analysis of Sec.~\ref{sec:static-spherical} and Eq.(\ref{phi-large-x}), we can see that
for $r \gg R_K$ and $r>R$ we have $\dd\phi/\dd x = y \sim 1/x^2 \ll 1$ and $K' \simeq 1$.
Therefore, at large distance beyond the K-mouflage radius (\ref{RK-def})
we find an increase of Newton's gravity by the constant multiplicative factor $1+2\beta^2$.
Within the K-mouflage radius $R_K$, where $y \gtrsim 1$, $K'$ becomes sensitive to the
nonlinear corrections associated with the nonstandard form of the kinetic term.
In particular, if $K'\gg 1$, as in the models (\ref{K0-no-sing}), the deviation from Newton's force
is suppressed by the factor $1/K'$ and we recover Newtonian gravity.
This nonlinear ``K-mouflage screening'' ensures the convergence to GR for small and dense
subgalactic and astrophysical systems and allows the models to satisfy observational constraints
from the Solar System or dwarf galaxies.
In particular, this means that for negative $\chi$ the derivative $K'$ must become large enough
to provide the required screening.
Thus, the screening criterion is
\beq
\mbox{screening:} \;\;\; r \ll R_K \;\;\; \mbox{and} \;\;\; K'(\chi) \gg 1  \;\; \mbox{for}
\;\; \chi \ll -1 .
\label{screening-crit}
\eeq

It is interesting to note that the constraint arising from Eq.(\ref{Geff-1}) is generically stronger
than the requirement of a unique well-defined scalar field profile for any matter density
profile, studied in Sec.~\ref{sec:static-spherical}.
Indeed, to obtain an efficient
screening we must have $K' > K'_{\rm obs}$ at large negative $\chi$,
where $K'_{\rm obs}$ is a lower positive bound derived from observations to ensure a small
enough deviation from GR on astrophysical scales.
Then, from the definition (\ref{Wy-def}) we obtain $W_{-} > K'_{\rm obs} y$ at large positive $y$,
which implies that $W_{-}(y)$ increases up to infinity when $y>0$.
This does not ensure that $W_{-}$ is monotonically increasing, as required for a unique well-defined
solution to Eq.(\ref {KG-W-1}) [a counter-example is provided by
$K'=K'_{\rm obs}+\alpha \sin^2(\omega\chi)$ with $\alpha>0$].
However, if $K'$ does not show oscillations but converges to a constant
$K'_{\infty} \geq K'_{\rm obs}$ or grows to $+\infty$ in a monotonic fashion, then $W_{-}$ is
monotonically increasing up to infinity.

Explicitly, for the models defined by (\ref{K-power-1}) with the constraints (\ref{K0-no-sing}),
where the scalar field profile is unique and well defined for any matter density profile,
we find that
\beqa
r>R \;\;  \mbox{and} \;\; r \ll R_K & : & \nonumber \\
&& \hspace{-3cm}  K' \sim ( |K_0| m)^{\frac{1}{2m-1}} \left( \frac{r}{R_K}
\right)^{\frac{-4(m-1)}{2m-1}} \gg 1
\label{Kp-small-r}
\eeqa
inside the K-mouflage  radius, and therefore the correction to Newton's law converges to zero well inside the K-mouflage radius (and outside of the central object).

\subsection{Screening of astrophysical and cosmological objects}
\label{sec:objects}

Screening of astrophysical objects can be easily identified by requiring that the
radius $R$ of the object is smaller than its K-mouflage radius $R_K$,
see also Eq.(\ref{screening-crit}).
From Eq.(\ref{RK-def}), we obtain at $z=0$,
\beq
\frac{R_K}{R}\approx \left( \frac{\beta R \delta}{R_{H_0}} \right)^{1/2} ,
\eeq
where $R_{H_0}=c/H_0$ is the Hubble radius and $1+\delta=\rho/\bar\rho_0$ is the
matter overdensity (we assume $\delta \gg 1$).
This gives the screening condition
\beq
\mbox{screening:} \;\; \frac{\beta R \delta}{R_{H0}} \gg 1 , \;\; \mbox{hence} \;\;
R \gg \frac{R_{H_0}}{\beta \delta} \;\; \mbox{or} \;\;
\delta \gg \frac{R_{H_0}}{\beta R} .
\label{screening-cond1}
\eeq
In the last two expressions, we expressed the screening condition as a lower bound
on the object radius, for a given density, or a lower bound on the density, for a given
radius. This criterion agrees with the criterion obtained in Eq.(23) of
\cite{Brax:2014aa} for large-scale cosmological structures.
More generally, this screening criterion is a condition on the product $(\rho R)$;
hence, $M/R^2 \sim |\nabla\Psi_{\rm N}|$, that is, the strength of the Newtonian
gravitational force, in agreement with the discussion in Sec.~II.C of
Ref.\cite{Brax:2014aa}.

Let us first consider typical astrophysical objects, such as stars, planets or asteroids,
with a density of the order of $\rho \sim 1\,{\rm g.cm}^{-3}$, hence
$\delta\sim 3.6 \times 10^{29}$. This yields $R \gg 0.035 \beta^{-1} \, \rm cm$.
Since we typically have $\beta \sim 0.1$, this means that all astrophysical objects
are far in the screened regime.
Moreover, dust grains throughout the Solar System,
such as Saturn rings, are screened by the Sun (i.e., ``blanket screening''
by a nearby massive object).
Indeed, the Solar System up to Neptun's orbit gives
$\delta \sim 2\times 10^{18}$ and
$R/R_{H_0} \sim 3.5 \times 10^{-14}$, hence
$\beta R \delta/R_{H0} \sim 7\times 10^4 \beta \gg 1$.

For the Milky Way, taking $R \sim 15$kpc and $M \sim 10^{12}M_{\odot}$,
we obtain $\beta R \delta/R_{H0} \sim 7 \beta \sim 1$.
This means that outer regions of the Galaxy, where $(\rho R)$ decreases as compared
with the value at $15$kpc, are unscreened, whereas inner regions, where
$(\rho R)$ increases somewhat, are screened.
Therefore, typical galaxies probe the transition between the screened and unscreened
regimes, whereas dwarfs should be mostly unscreened.
Thus, galaxies are promising objects to constrain such K-mouflage models.

Finally, let us consider clusters of galaxies. With $R \sim 1$Mpc and $\delta \sim 200$,
we obtain $\beta R \delta/R_{H0} \sim 0.05 \beta \ll 1$, which means that outer
regions down to the virial radius are unscreened.
This agrees with the analysis of Ref.\cite{Brax:2014ab}, where we noticed that
large-scale cosmological structures are unscreened and in the linear regime
for the scalar field sector [i.e., even though the matter density contrast can be in the
mildly nonlinear regime, the Klein-Gordon equation (\ref{KG-2}) can be linearized
over $\varphi$].
However, cluster cores would probe the screened regime and the nonlinear
part of the kinetic function $K(\chi)$.

\subsection{Spherical waves}
\label{sec:waves}

\subsubsection{Linear stability}
\label{sec:linear-stability}

We can now study the dynamics of scalar field perturbations on the spherical background
obtained in Sec.~\ref{sec:screening}.
When we consider radii outside of the object, that is, in the vacuum (i.e., the matter density
is zero throughout space at $r>R$, but the background scalar field $\varphi$ is
not zero), we can study local scalar field perturbations at fixed vanishing matter density
(hence there are no coupled local matter density perturbations).
We focus on spherically symmetric fluctuations and we investigate their linear stability.
Thus, writing the scalar field as
\beq
\varphi(r,t) = \bar\varphi(r) + \delta\varphi(r,t) ,
\label{dphi-def}
\eeq
where $\bar\varphi$ is the static spherical solution of Eq.(\ref{KG-4}) obtained
in Sec.~\ref{sec:static-spherical}, and linearizing the Klein-Gordon equation (\ref{KG-3}),
we obtain
\beq
\bar{K}' \frac{\pl^2 \delta\varphi}{\pl t^2} - \frac{c^2}{r^2} \frac{\pl}{\pl r}
\left[ r^2 (\bar{K}' + 2 \bar{\chi} \bar{K}'' ) \frac{\pl \delta\varphi}{\pl r} \right] = 0 ,
\label{wave-1}
\eeq
where $\bar\chi$ and $\bar{K}$ are the spherical background quantities.
This is a hyperbolic partial differential equation, which reduces for short wavelengths to the
wave equation
\beq
\frac{\pl^2\delta\varphi}{\pl t^2} - c_s^2 \, \frac{\pl^2\delta\varphi}{\pl r^2} = 0 ,
\label{wave-2}
\eeq
with the position-dependent propagation speed
\beq
c_s^2 = \frac{\bar{K}'+2\bar{\chi}\bar{K}''}{\bar{K}'} \; c^2 = \frac{\bar{W}_{-}'}{\bar{K}'} \; c^2
= \frac{\bar{y} \bar{W}_{-}'}{\bar{W}_{-}} \; c^2 ,
\label{c-speed-1}
\eeq
where $\bar{W}_{-}' = \frac{\dd W_{-}}{\dd y}(\bar{y})$ and we have used the definition
(\ref{Wy-def}).
The ratio $c_s/c$ is formally the inverse of the one obtained in the uniform time-dependent
cosmological background \cite{Brax:2014aa}. However, the background values $\bar{K}'$
and $\bar\chi$ are different and bear no relations.
Indeed, in the cosmological context we only probe the part $\chi>0$ of the kinetic function
$K(\chi)$ whereas in the regime studied in this paper, and in Eq.(\ref{c-speed-1}),
we have $\chi<0$.

For the models such as (\ref{K0-no-sing}), where $W_{-}(y)$ is monotonically increasing up to
$+\infty$ over $[0,y_-[$ and there is a unique well-defined scalar field profile,
we have $\bar{W}_{-}'>0$ and $\bar{K}'>0$. Therefore, $c_s^2$ is positive and Eq.(\ref{wave-2})
gives rise to traveling waves. Thus there are no spherical instabilities at the linear level.
More generally, the condition for linear stability is automatically satisfied once the conditions for
a well-defined static profile are verified, see Eqs.(\ref{condition-static-W}) and
(\ref{condition-static-K}):
\beq
\mbox{linear stability:} \;\;\; K' > 0, \;\;\; W_{-}' \equiv K' + 2 \chi K'' \geq 0 ,
\label{condition-linear-stability}
\eeq
over the range $\chi_-<\chi\leq 0$, $0\leq y < y_-$.

\subsubsection{Superluminality}
\label{sec:superluminality}

At large radius $r\rightarrow \infty$, where we have $\bar\chi \rightarrow 0$ and the expansion
(\ref{K-chi=0}), we obtain $c_s \rightarrow c$.
At finite radius, since $\bar{\chi} < 0$, the propagation speed is smaller than the speed of
light if $\bar{K}''>0$, and greater if $\bar{K}''<0$.
For the explicit models (\ref{K0-no-sing}) we have $\bar{K}''<0$ hence $c_s > c$.
More generally, if we consider non-polynomial functions $K(\chi)$ with a power-law
behavior at large negative $\chi$, we have
\beq
\chi \rightarrow -\infty: \;\;\; K \sim - | \chi |^m , \;\;\;  c_s^2 \sim (2m-1) c^2 ,
\label{chi-m-infty}
\eeq
where $m$ is not necessarily an integer. Then, the requirement for small-scale high-density
screening (\ref{Geff-1}) implies $m \geq 1$, which also ensures that $W_{-}(y) \sim y^{2m-1}$
is monotonically increasing up to $+\infty$, and $c_s \geq c$.
Therefore, simple K-mouflage models that can accommodate arbitrary matter density
profiles and display nonlinear screening have propagation speeds greater than the speed of light
for scalar field perturbations, see also \cite{Burrage2014}.

This is actually a generic feature of K-mouflage models.
Indeed, Eq.(\ref{c-speed-1}) also reads as $c_s^2 = 1 +2 \bar{\chi} \bar{K}''/\bar{K}'$, and since
$\bar\chi<0$ in the quasistatic regime, a propagation speed smaller than the speed of light requires
that $\bar{K}''$ and $\bar{K}'$ be of the same sign.
In models where $K'(\chi)>0$ (to avoid ghosts in the cosmological regime, $\chi>0$, and to avoid divergences and instabilities
at the point where $K'$ would go through zero to change sign), this implies $K''>0$ over $\chi<0$.
Then, $K'(\chi)$ is a monotonically increasing function on the negative real axis, whence
$0< K'(\chi) < K'(0)=1$ for $\chi<0$. This prevents any nonlinear K-mouflage screening, which
relies on the condition $|K'| \gg 1$ in small-scale high-density environments, see Eq.(\ref{Geff-1}).

Therefore, realistic models with efficient screening must show superluminality.
Of course, it is always possible to have $c_s < c$ over some limited range, but this
cannot hold for all regimes $\chi<0$.

\section{Generalized solutions}
\label{sec:generalized}

\subsection{Cases where $W_{-}(y)$ cannot be inverted}
\label{sec:not-inverted}

As described in Sec.~\ref{sec:static-spherical}, when the function $W_{-}(y)$ defined by
Eq.(\ref{Wy-def}) is monotonically increasing up to $+\infty$ over $0\leq y < y_-$,
where $y_-$ can be finite or $+\infty$, Eq.(\ref{KG-W-1}) can be inverted and we obtain
a well-defined static profile for the scalar field for any matter density profile.

This is no longer possible when $W_{-}(y)$ is not monotonically increasing up to $+\infty$.
Two different cases can be encountered, depending on whether $W_{-}(y)$ is bounded or not.

\subsubsection{No solution when $W_{-}(y)$ is bounded}
\label{sec:bounded}

The first case where $W_{-}(y)$ is not monotonically increasing up to $+\infty$ corresponds to
functions $W_{-}(y)$ that are bounded, with $|W_{-}(y)| \leq W_{\rm max}$ for all $y \geq 0$
[because $W_{-}(y)$ is an odd function of $y$, it is sufficient to consider the range $y \geq 0$].
Then, for high-density objects where $m(x)/x^2$ can reach values beyond $W_{\rm max}$,
no solution can be found to Eq.(\ref{KG-W-1}) and there exists no static scalar field profile
that is valid throughout space.
However, at small and large radii, where the Newtonian gravitational force is small enough
[i.e., $m(x)/x^2$ is below the maximum $W_{\rm max}$], one can still define a local static profile,
that is, a local solution to  Eq.(\ref{KG-W-1}).

Thus, kinetic functions $K(\chi)$ that enter this class cannot provide realistic models,
or they are incomplete and one must add higher-order corrections that ensure a better behaved
$W_{-}(y)$, so as to provide a well-defined static profile for the scalar field for any matter density profile and at all radii.

\subsubsection{Infinite number of discontinuous solutions when $W_{-}(y)$ is not bounded and non-monotonic}
\label{sec:discontinuous}

The second case where $W_{-}(y)$ is not monotonically increasing up to $+\infty$ corresponds to
functions $W_{-}(y)$ that are not bounded, so that at each radius one can always find at least
one solution $y(x)$ to Eq.(\ref{KG-W-1}), but $W_{-}(y)$ is not monotonically increasing over
$y \geq 0$.
Then, for high-density objects, starting from $x\rightarrow +\infty$ and $y \simeq 1/x^2$,
as we move closer to the object and $m(x)/x^2$ increases, we meet the first local maximum
$y_{\rm max}^{(1)}$ of $W_{-}(y)$ at some point $x_1$.
We can extend the solution $y(x)$ to smaller radii by allowing for discontinuous solutions
$y(x)$, with $y(x_1^+) = y_{\rm max}^{(1)}$ and by jumping at $x_1^-$ to a point $y$
that is at a finite distance from $y_{\rm max}^{(1)}$, within an interval where the function
$W_{-}(y)$ runs from $W_{\rm min}^{(2)}$ to $W_{\rm max}^{(2)}$ with
$W_{\rm min}^{(2)} < W_{\rm max}^{(1)} < W_{\rm max}^{(2)}$.
If there are several local maxima, we can build a solution with several jumps.

If $W_{-}(y) \rightarrow -\infty$ on the positive axis $y \geq 0$, we eventually need to jump to the
negative axis $y<0$ to find an interval where $W_{-}(y) \rightarrow +\infty$, which provides
a solution for any matter density profile.
A negative $y$ also implies a negative $K'$ from Eq.(\ref{Wy-def}), because $W_{-}=m(x)/x^2$
is always strictly positive. This means that the fifth force decreases Newtonian gravity,
that is, the effective Newton's constant is smaller than $\cG$ (or negative) from Eq.(\ref{Geff-1}).

These solutions are not unique and we can actually build an infinite number of them.
Indeed, instead of jumping at position $x_1$ when we reach the local maximum
$W_{\rm max}^{(1)}$, we could have chosen to make the jump at any slightly larger
radius $x \gtrsim x_1$, such that $W_{-}[y(x)] > W_{\rm min}^{(2)}$.

To ensure the linear stability of such solutions with respect to radial perturbations,
we can see from Eq.(\ref{c-speed-1}) that we must always have $yW_{-}' \geq 0$.
This implies, for instance, that as we patch together several intervals on the
positive real axis $y \geq 0$ to build a discontinuous profile, we must only use
intervals where $W_{-}(y)$ is increasing, with $W_{-}' \geq 0$.
In a similar fashion, intervals on the negative real axis must satisfy $W_{-}' \leq 0$.

\subsection{Steady-state solutions}
\label{sec:steady}

In the previous sections we considered the static solutions of the Klein-Gordon equation
(\ref{KG-3}), given by Eqs.(\ref{KG-static-1}) or (\ref{KG-4}). However, for a static density
profile it is possible to find more general ``steady-state'' solutions for the scalar field,
of the form:
\beq
\varphi(\vr,t) = \nu t + \hat{\varphi}(\vr) ,
\label{nu-def}
\eeq
where the time-independent function $\hat{\varphi}$ obeys
\beq
\nabla_{\vr} \cdot (\nabla_{\vr} \hat{\varphi} \; K' ) =  \frac{\beta\rho}{c^2 M_{\rm Pl}} \;\;\;
\mbox{with} \;\;\; \chi= \frac{\nu^2-c^2 (\nabla_{\vr}\hat{\varphi})^2}{2{\cal M}^4} .
\label{KG-steady}
\eeq
Thus, the scalar field shows a linear time dependence, with a space-independent pre-factor,
in addition to the time-independent but space-dependent part $\hat{\varphi}$.
Because the kinetic variable $\chi$ in Eq.(\ref{chi-r}) only involves first-order derivatives,
the Klein-Gordon equation (\ref{KG-steady}) does not show any time dependence, even though
the field $\varphi$ includes a linear time-dependent term.
Moreover, the fifth force, which is set by $-\nabla_{\vr} \varphi$ as in Eqs.(\ref{Euler2}) and
(\ref{5th-force}), is also time-independent.

Then, for a given static density profile $\rho(r)$ we can build an infinite number of steady solutions,
parametrized by the constant $\nu$. In particular, in the cases where the function $W_{-}(y)$
defined in Eq.(\ref{Wy-def}) is bounded and we could find no static solution in
Sec.~\ref{sec:bounded},
we can now find an infinite number of solutions by choosing large
values of $\nu^2$, if the derivative of the kinetic function $K'(\chi)$ goes to $+\infty$ on the
positive semi-axis $\chi>0$, which is required to obtain a realistic cosmology up to high
redshifts \cite{Brax:2014aa,Brax:2014ab}.

To remove any ambiguity, we must note that on large scales, within the cosmological setting,
the scalar field is actually time-dependent as it follows the evolution of the cosmological background.
More precisely, at the background level the scalar field $\bar\varphi$ is the solution of the evolution
equation \cite{Brax:2014aa}
\beq
\mbox{cosm. background:} \;\;\;
a^3 \dot{\bar\varphi} \bar{K}' = - \int_0^t \dd t' \, \bar\rho_0 \frac{\dd \bar{A}}{\dd\bar{\varphi}}(t') ,
\label{phi-backgd1}
\eeq
which gives
\beq
\bar{\varphi} \sim - \frac{\beta\bar{\rho}t^2}{M_{\rm Pl} \bar{K}'}  ,  \;\;\;
\bar{\nu} \equiv \dot{\bar\varphi} \sim - \frac{\beta\bar{\rho} t}{M_{\rm Pl} \bar{K}'} ,
\label{nu-bar-def}
\eeq
where we defined $\bar\nu$ as the time-derivative of the cosmological background value of the
scalar field, and $t$ is the age of the Universe at the redshift of interest.
This uniform cosmological time dependence, which we actually neglected in this paper, must be
distinguished from the factor $\nu t$ in the generalized solutions (\ref{nu-def}), which would arise
from the small-scale nonlinearities.
Indeed, the latter should be understood as a hypothetical feedback from small nonlinear scales
up to cosmological scales of order $c t$, the propagation speed of scalar field fluctuations being
of order $c$.
Thus, from Eq.(\ref{KG-steady}) the coefficient $\nu$ that would arise from small-scale nonlinearities
would be of order
\beq
\nu \sim \frac{c \hat{\varphi}}{r} \sim \frac{\beta \rho r}{c M_{\rm Pl} K'} ,
\label{nu-small-scale}
\eeq
whence
\beq
\frac{\nu}{\bar{\nu}} \sim \frac{\rho}{\bar\rho} \, \frac{r}{c t_0} .
\label{nu-nubar}
\ee
For the Solar System, up to the Jupiter orbit, we have $\rho/\bar\rho \sim 10^{20}$ and
$r/(c t_0) \sim 10^{-14}$, whence $\nu/\bar\nu \sim 10^6$. Thus, the cosmological value
$\bar\nu$ is indeed negligible as compared with the small-scale value $\nu$ that is required
to build a truly generalized solution (\ref{nu-def}), where the time-dependent term $\nu t$
cannot be neglected as compared with the static term $\hat\varphi$.

Because the constant $\nu$ in Eq.(\ref{nu-def}) cannot depend on space the time-dependent part
$\nu t$ extends over all space. More precisely, in the cosmological setting, it extends up to
scales of order $c t$, over which the system has had time to relax, where $t$ is the age of the
Universe. Indeed, if we consider several nonlinear matter overdensities, separated by large
distances, each one being at the center of domains $V_i$ characterized by different values $\nu_i$,
we find below from Eq.(\ref{shock-1}) that the discontinuity fronts between these regions
cannot be motionless.
Then, this would no longer yield a time-independent fifth force, because of the Dirac component along
the moving domain boundaries.
However, one would expect the system to relax towards a unique value $\nu$ within a cosmological
domain of size $c t$, which would cover the observed Universe.

Fortunately, our numerical analysis, presented in Sec.~\ref{sec:relax-DBI-} below, shows that
such solutions are not achieved, even in the case where there is no static solution throughout
all space because the function $W_-(y)$ is bounded.
We will find that in such cases the system relaxes to the static solution at larger distance
from high-density regions, where it is well defined, while shocks and ever-increasing gradients
are confined to small scales close to the objects.
Thus, even in these somewhat pathological cases, the small-scale nonlinearities and singularities
do not propagate to large scales and have no impact on the cosmological behavior.

\section{Dynamics and Relaxation of the scalar field profile}
\label{sec:dynamics}

\subsection{Characteristics}
\label{sec:characteristics}

To check whether the solution with $\nu=0$ is indeed reached by the system, and what
the obtained behavior is  in cases where there is no well-defined static solution [i.e., $W_{-}(y)$
is not monotonically increasing up to $+\infty$], we consider in this section the relaxation of the
scalar field profile.
To study the evolution with time of the scalar field, within a given matter density
background (that may depend on time), we must solve the Klein-Gordon equation (\ref{KG-3}).
For a spherically symmetric density profile this also writes as
\beq
\frac{\pl}{\pl t} \left( \frac{\pl \varphi}{\pl t} \; K' \right)
- \frac{c^2}{r^2} \frac{\pl}{\pl r}  \left( r^2 \frac{\pl\varphi}{\pl r} \; K' \right) =
- \frac{\beta\rho}{M_{\rm Pl}} ,
\label{KG-t-1}
\eeq
with
\beq
\chi = \frac{1}{2{\cal M}^4} \left[ \left( \frac{\pl \varphi}{\pl t} \right)^2
 - c^2 \left( \frac{\pl\varphi}{\pl r} \right)^2 \right] .
\label{chi-t-1}
\eeq
For an object of mass $M$, making the same changes of variables as in Eq.(\ref{KG-4})
in terms of the K-mouflage radius $R_K$ defined in Eq.(\ref{RK-def}) and the
scalar field normalization $\varphi_K$ of Eq.(\ref{phiK-def}),
with $x=r/R_K$, $\phi(x)= \varphi(r)/\varphi_K$, and $\tau= c t/R_K$, Eq.(\ref{KG-t-1}) takes
the dimensionless form
\beq
\frac{\pl}{\pl\tau} \left( \frac{\pl\phi}{\pl\tau} K' \right) - \frac{1}{x^2} \frac{\pl}{\pl x}
\left( x^2 \frac{\pl\phi}{\pl x} K' \right) = - \eta ,
\label{KG-t-2}
\eeq
with 
\beq
\eta(x,\tau)= \rho(r,t) \; \frac{4\pi R_K^3}{M} .
\label{eta-def}
\eeq
This is a quasilinear second-order equation because by expanding all terms we can see
that the second-order derivatives only appear linearly in the partial differential equation.

It is convenient to transform the second-order equation (\ref{KG-t-2}) as a system of two
first-order equations, by introducing
\be
u = \frac{\pl\phi}{\pl\tau} , \;\;\; v= \frac{\pl\phi}{\pl x} , \;\;\; \chi= \frac{u^2-v^2}{2} ,
\label{u-v-def}
\eeq
which gives
\beqa
\frac{\pl}{\pl\tau} ( x^2 u K' ) + \frac{\pl}{\pl x} ( m - x^2 v K' ) & = & 0 ,
\label{u-t-1} \\
\frac{\pl v}{\pl\tau} - \frac{\pl u}{\pl x} & = & 0 ,
\label{v-t-1}
\eeqa
where $m(x,\tau)= \int_0^x \dd x' \, x'^2 \eta(x',\tau) = M(<r,t)/M$ is the mass within the radius 
$r$ normalized as in Eq.(\ref{KG-5}).
Here we have written the system (\ref{u-t-1})-(\ref{v-t-1}) in a flux-conservative form, with
the conserved densities $\{x^2uK',v\}$ and the fluxes $\{m-x^2vK',-u\}$.
Such forms are better suited to numerical computations.

The expanded form reads as the quasi-linear first-order system
\beqa
\hspace{0cm} (K' \!+\! u^2K'') \frac{\pl u}{\pl\tau} - K'' u v \left( \frac{\pl u}{\pl x} \!+\! \frac{\pl v}{\pl\tau} \right)
- (K' \!-\! v^2K'') \frac{\pl v}{\pl x}  && \nonumber \\
&& \hspace{-3cm} = \frac{2}{x} v K' - \eta  ,
\label{u-t-2} \\
&& \hspace{-8cm} \frac{\pl v}{\pl\tau} - \frac{\pl u}{\pl x} = 0 .
\label{v-t-2}
\eeqa
This system can be analyzed by the method of characteristics.
The two trajectory characteristics are
\beq
\left(\frac{\dd x}{\dd\tau} \right)_{\pm} = c_{\pm} \equiv \frac{-K'' u v \pm \sqrt{K' [K'+K'' (u^2-v^2)]}}{K'+u^2K''} ,
\label{cpm-def}
\eeq
and we have the two evolution equations along both sets of characteristics
\beq
\dd u + c_{\mp} \dd v = \frac{(2/x) v K' - \eta}{K'+u^2K''}  \, \dd t \;\;\; \mbox{along} \;\;\;
\dd x = c_{\pm} \dd t .
\label{carac-2}
\eeq
Equations (\ref{u-t-2})-(\ref{v-t-2}) form a first-order hyperbolic system when the characteristic
speeds $c_{\pm}$ are real and an elliptic system when they are complex.
Therefore, to obtain a well-defined Cauchy problem, that is, the evolution of the scalar field can
be obtained from an initial condition by solving these differential equations forward in time,
the argument of the square root in Eq.(\ref{cpm-def}) must always remain positive.
If this is not the case, that is, at some time we obtain a negative argument over some interval
in $x$, we can no longer solve the evolution of the scalar field at later times as we obtain
an elliptic problem that requires boundary conditions over all boundaries (unless this domain closes at a later finite time and we can solve the left and right boundaries from the hyperbolic domains).
For the standard kinetic term, where $K'=1$ and $K''=0$, the characteristic speeds are
constant with $c_{\pm}=\pm 1$ and we obtain a simple Cauchy problem.

If we start from the static profiles described in Sec.~\ref{sec:static-spherical},
given by Eq.(\ref{KG-W-1}), with a zero time-derivative, we have $u=0$, $v=y$,
and $W_{-}'=K'-y^2 K''$ [where we again note $W_{-}'(y)=\dd W_{-}/\dd y$], so that the
characteristic speeds are
\beq
\mbox{on static profile:} \;\;\; c_{\pm} = \pm \sqrt{\frac{W_{-}'}{K'}} ,
\label{cpm-static}
\eeq
where we used the property $K'>0$ (to ensure a well-defined static profile and to avoid
a divergent fifth force, see Sec.~\ref{sec:screening}).
Since we also have $W_{-}' \geq 0$ we obtain a well-defined Cauchy problem for the evolution
close to this static profile.
Moreover, we obtain $c_{\pm}^2 = c_s^2$, where $c_s$ defined in
Eq.(\ref{c-speed-1}) is the propagation speed of radial waves around the static profile.

More generally, to ensure the hyperbolicity of the differential equation (\ref{KG-t-2}) for
any initial condition and at any time, we must have
$K' ( K'+2\chi K'' ) \geq 0$ for all values of $\chi$, both on the negative and positive semi-axis.
If we consider models where $K' \geq 0$, to avoid ghosts in the cosmological regime $\chi>0$,
and to ensure a well-defined profile in the static regime $\chi<0$, this condition
reads as $K'+2\chi K'' \geq 0$ over the whole real axis for $\chi$.
This is in fact the only possibility. Indeed, if we only require that the product $K'(K'+2\chi K'')$
remains positive, starting from $K'=1$ at $\chi=0$, we can see that $K'$ and $K'+2\chi K''$
must vanish at the same point $\chi_0$ if $K'$ can change sign.
Then, writing $K' \sim K'_0 (\chi-\chi_0)^n$ around $\chi_0$, with $n \geq 2$, we have
$K' ( K'+2\chi K'') \simeq K_0^{'2} 2 n \chi_0 (\chi-\chi_0)^{2n-1}$ which changes sign at $\chi_0$
and does not satisfy the positivity constraint.
Therefore, the hyperbolicity condition reads as
\beq
\mbox{hyperbolicity:} \;\;\; K' \geq 0 , \;\;\; K'+2\chi K'' \geq 0 ,
\label{hyperbolicity}
\eeq
for all values of $\chi$, from $\chi_-$ to $\chi_+$
where $\chi_{\pm}$ may be finite or $\pm\infty$, depending on the domain of definition of
$K(\chi)$.
As could be expected, this is a stronger constraint than the condition for a well-defined static
profile and stable fluctuations, which only read as $K'+2\chi K'' \geq 0$ over the negative real axis
$\chi_-<\chi<0$, see Eqs.(\ref{condition-static-K}) and (\ref{condition-linear-stability}).

For nonstandard kinetic functions $K(\chi)$ the characteristic speeds $c_{\pm}$ are not
constants and two different characteristics of the same set may cross, leading to the formation
of shocks. Then, from the flux-conservative form (\ref{u-t-1})-(\ref{v-t-1}), across a shock
at position $x_s(\tau)$ moving at speed $\dd x_s/\dd\tau$, we have the discontinuity conditions
\beq
[ v K' ] = - \frac{\dd x_s}{\dd\tau} [ u K' ] , \;\;\;
[ u ] = - \frac{\dd x_s}{\dd\tau} [ v ] ,
\label{shock-1}
\eeq
which implies
\beq
\frac{[v K' ]}{[u K']} = \frac{[u]}{[v]} ,
\label{shock-2}
\eeq
where we noted $[X]= X(x_s^+)-X(x_s^-)$ the discontinuity of a quantity $X$ through the shock.
In particular, if the shock is motionless $u$ and $vK'$ are continuous.
The continuity of $u$ through motionless shocks implies that several steady-state solutions of the
form (\ref{nu-def}) with different values of $\nu$ cannot be patched over time-independent domains.

\subsection{Numerical analysis}
\label{sec:numerical}

\subsubsection{Matter density background}
\label{sec:matter-background}

As numerical examples of the relaxation of the scalar field within a given matter density
background, we consider the static Gaussian matter density profiles
\beq
\rho_{\rm static}(r) = \rho_0 \; e^{-r^2/R^2} ,
\label{rho-Gauss}
\eeq
which read in terms of the dimensionless variables $x$ and $\eta$, as
\beq
x= \frac{r}{R_K} , \;\;\;
\eta_{\rm static}(x) = \frac{4}{\sqrt{\pi}} \left( \frac{R_K}{R} \right)^3 e^{-(x R_K/R)^2} .
\label{eta-Gauss}
\eeq
Here, $R$ is the characteristic radius of the overdensity. For a large ratio $R/R_K \gg 1$,
the dimensionless overdensity has a small amplitude $\eta_{\rm static}(0) \ll 1$ and a wide
characteristic size $x \sim R/R_K \gg 1$, and we are in the unscreened weak-field regime,
with $| \chi | \ll 1$ and $K' \simeq 1$.
For a small ratio $R/R_K \ll 1$, we have a high amplitude $\eta_{\rm static}(0) \gg 1$ and small size
$R/R_K \ll 1$, and we are in the screened strong-field regime, with $| \chi | \gg 1$
and $K' \gg 1$.

We can study the relaxation of the scalar field in this fixed matter density background, starting
with a uniform scalar field configuration, with $u=v=0$ at $\tau=0$.
However, in realistic cases the matter density fluctuations are not built instantaneously but
on a long formation time-scale $t_{\rm f}$ that gives time to the scalar field to follow the matter
evolution through a series of quasistatic states.
To describe this more realistic configuration, we define the matter density background to reach
the final static state (\ref{rho-Gauss})-(\ref{eta-Gauss}) only after a time $t_{\rm f}$, and at ealier
times we write
\beqa
t < t_{\rm f} & : & \rho(r,t) = \frac{t}{t_{\rm f}} \, \rho_{\rm static}(r) , \label{rho-t}  \\
\tau < \tau_{\rm f} & : & \eta(x,\tau) = \frac{\tau}{\tau_{\rm f}} \, \eta_{\rm static}(x) . \label{eta-tau}
\eeqa
The formation time $t_{\rm f}$ cannot be arbitrarily small because matter cannot inflow from larger scales
at arbitrarily large velocities, to build the central overdensity. Defining the typical matter velocity
as $v_{\rm f} = \alpha_{\rm f} c$, with $\alpha_{\rm f} < 1$, and the typical formation time as
$t_{\rm f} = R/v_{\rm f}$, we have
\beq
\tau_{\rm f} = \frac{1}{\alpha_{\rm f}} \, \frac{R}{R_K} = \frac{\tau_{\phi}}{\alpha_{\rm f}} ,
\;\;\; \mbox{with} \;\;\; \tau_{\phi} = \frac{R}{R_K} .
\label{tauf-def}
\eeq
Here $\tau_{\phi}$ is the typical time scale for the scalar field relaxation, that we estimate
from Eq.(\ref{KG-t-2}) as $\tau_{\phi} \sim x \sim R/R_K$.
The simplified model (\ref{eta-tau})-(\ref{tauf-def}) describes the formation of a local overdensity
as matter flows inward from a low-density background, on the time-scale $t_{\rm f}$.
For small values of $\alpha_{\rm f}$, $\tau_{\rm f}$ becomes very large and the scalar field
has the time to follow the growth of the matter overdensity through a sequence of quasistatic
states.
In the following figures we consider the conservative case $\alpha_{\rm f}=0.1$, but for the formation
of galaxies or clusters of galaxies, where velocities are on the order of a few hundred km/s,
we would have $\alpha_{\rm f} \sim 10^{-3}$.

As can be read from Eqs.(\ref{u-t-1})-(\ref{v-t-1}), in agreement with Eq.(\ref{KG-5}),
the static scalar field profile $\phi_{\rm static}(x)$ associated with a given matter density
profile is given by:
\beq
\mbox{static:} \;\;\; u=0 , \;\;\; v K'(-v^2/2) = m(x)/x^2 .
\label{static-u-v-x}
\eeq
At times $\tau \gg \tau_{\rm f}$, long after the matter density profile has reached the static configuration
(\ref{eta-Gauss}), we expect the scalar field to have relaxed to the static solution (\ref{static-u-v-x}).
For small enough $\alpha_{\rm f}$, that is, large formation time-scales, we also expect the scalar field
to follow the evolution of the matter profile by going through the sequence of static solutions
(\ref{static-u-v-x}), where $m(x,\tau)$ slowly evolves with time.

\subsubsection{Relaxation when $W_{-}(y)$ monotonically increases to $+\infty$ over $0\leq y <+\infty$}
\label{sec:relax-monotonic}

\begin{figure*}
\begin{center}
\epsfxsize=8.5 cm \epsfysize=5.8 cm {\epsfbox{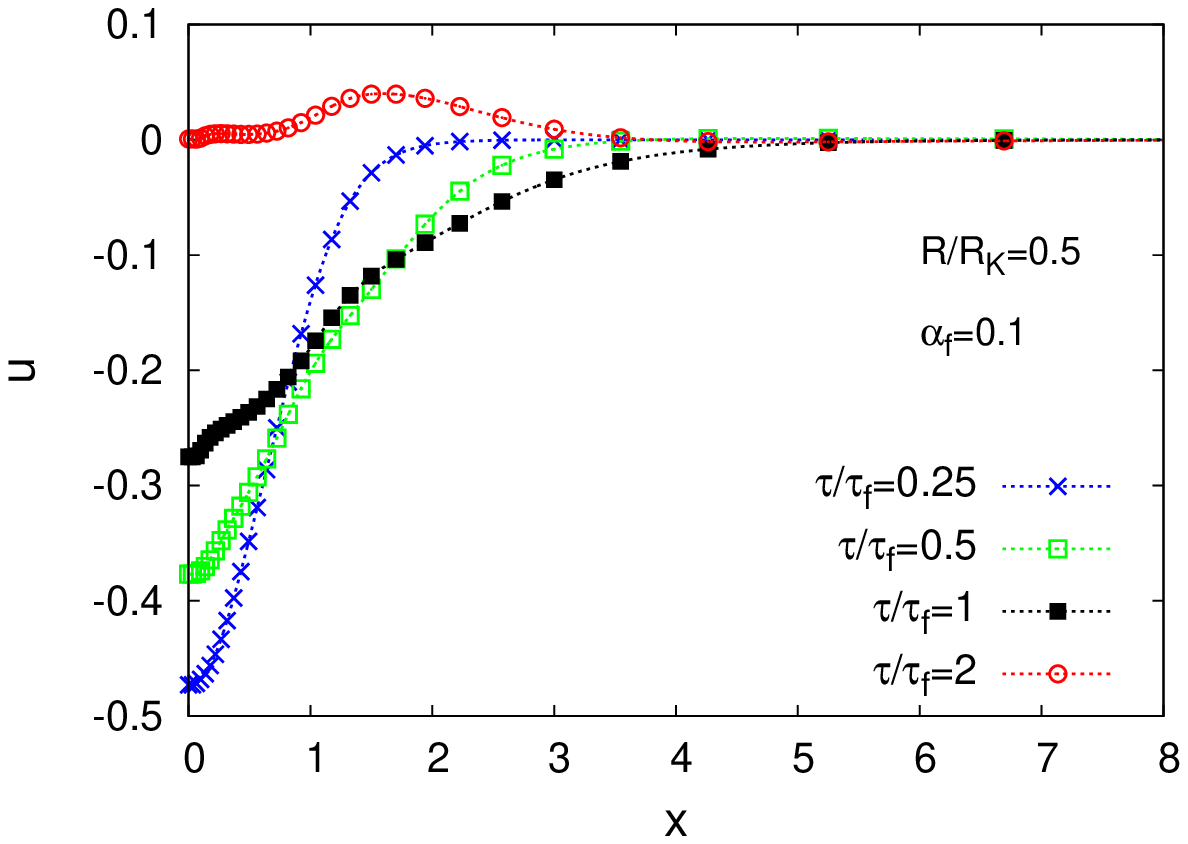}}
\epsfxsize=8.5 cm \epsfysize=5.8 cm {\epsfbox{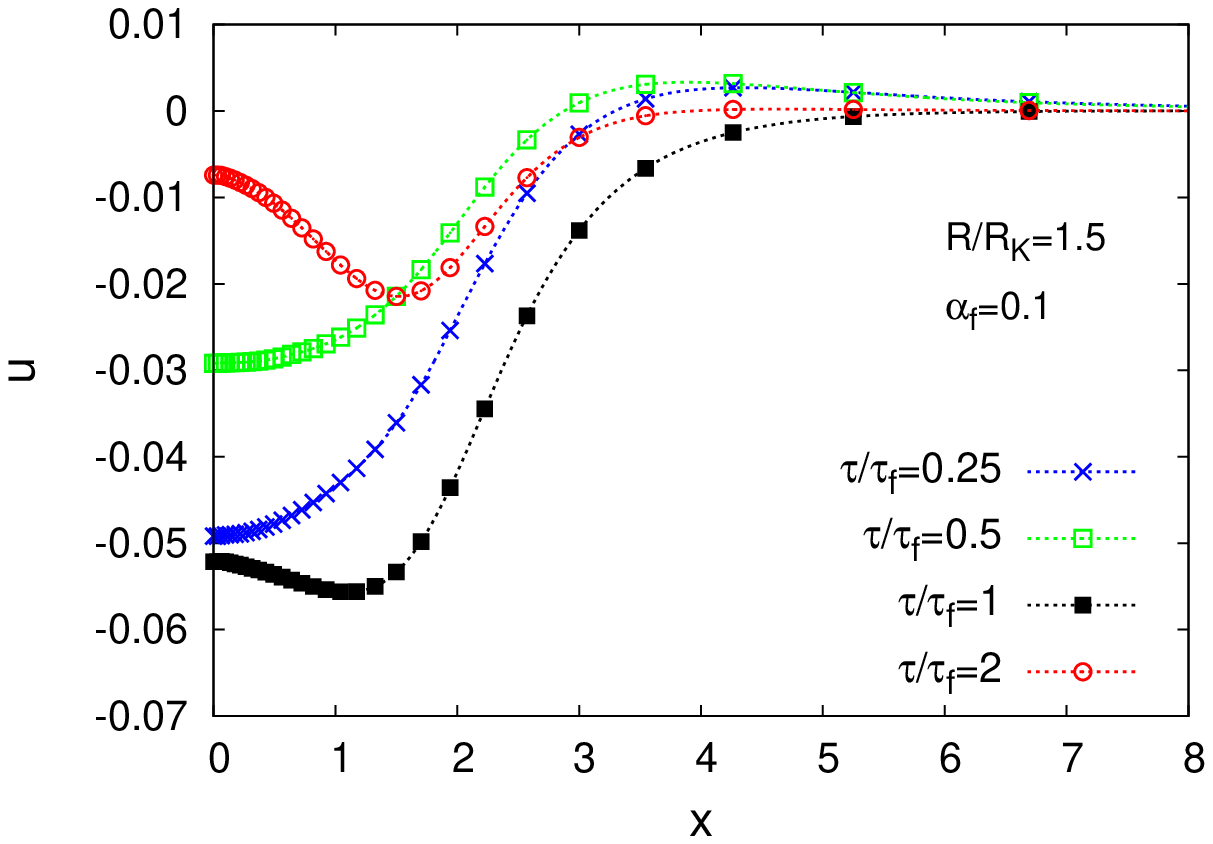}}\\
\epsfxsize=8.5 cm \epsfysize=5.8 cm {\epsfbox{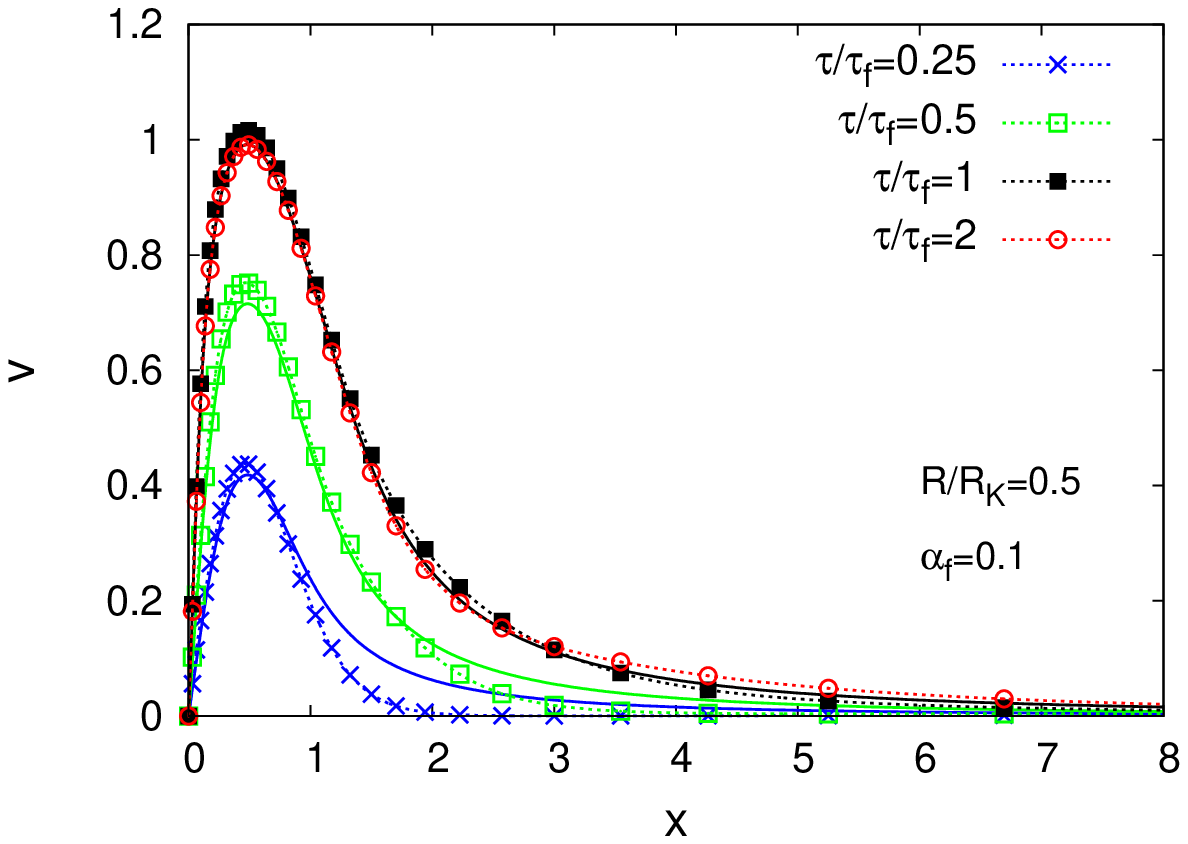}}
\epsfxsize=8.5 cm \epsfysize=5.8 cm {\epsfbox{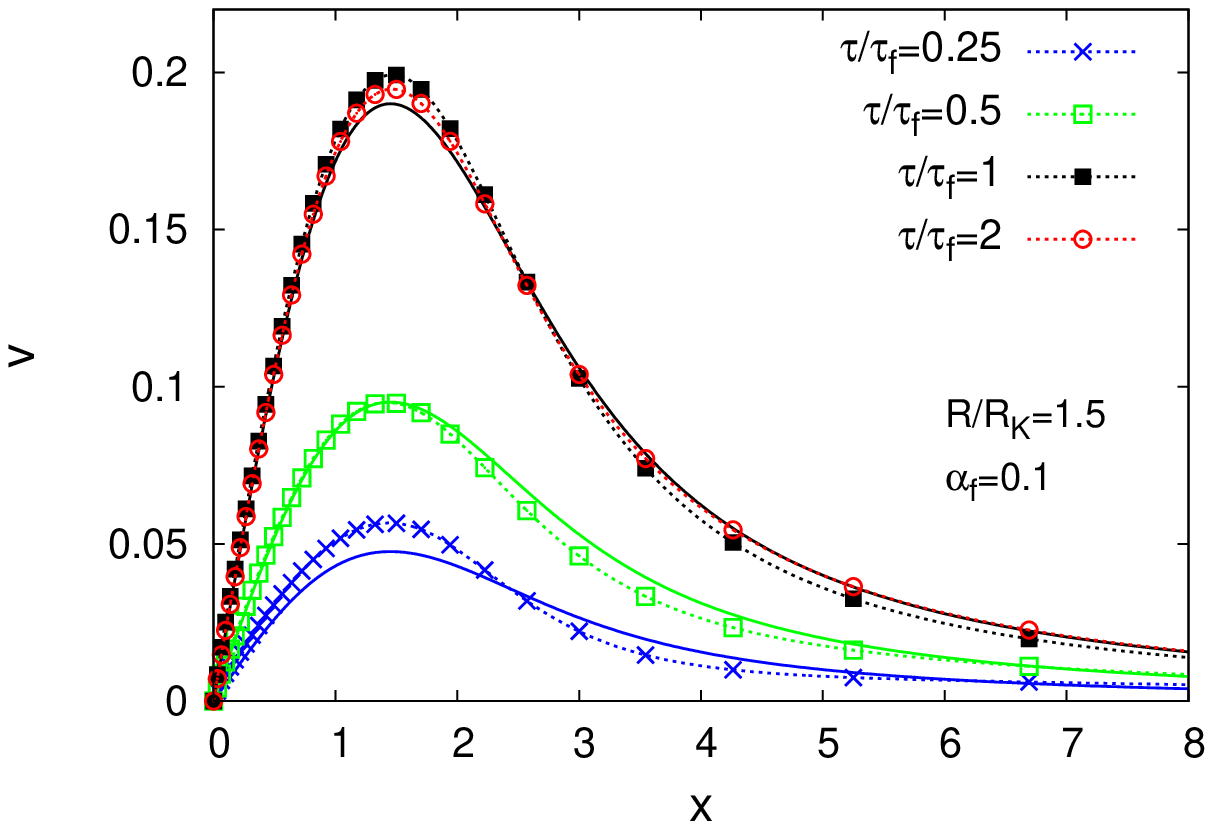}}\\
\epsfxsize=8.5 cm \epsfysize=5.8 cm {\epsfbox{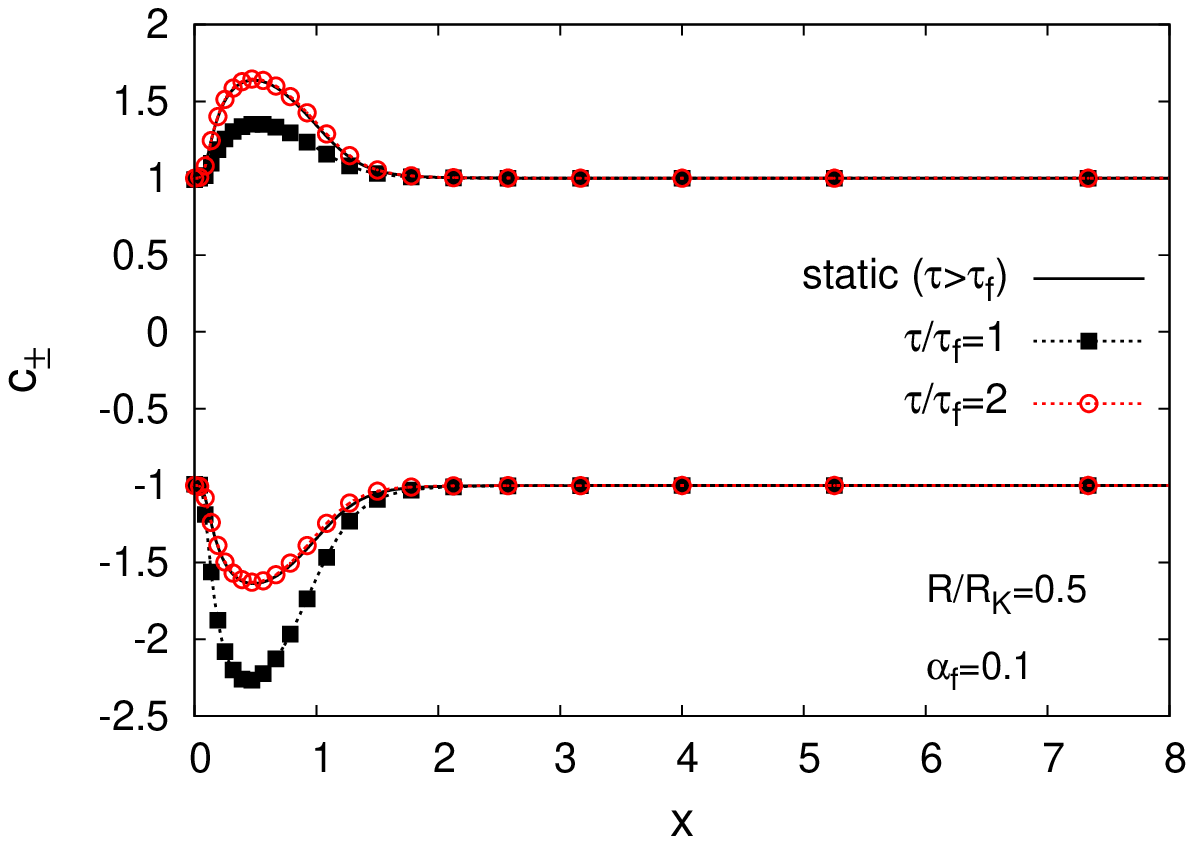}}
\epsfxsize=8.5 cm \epsfysize=5.8 cm {\epsfbox{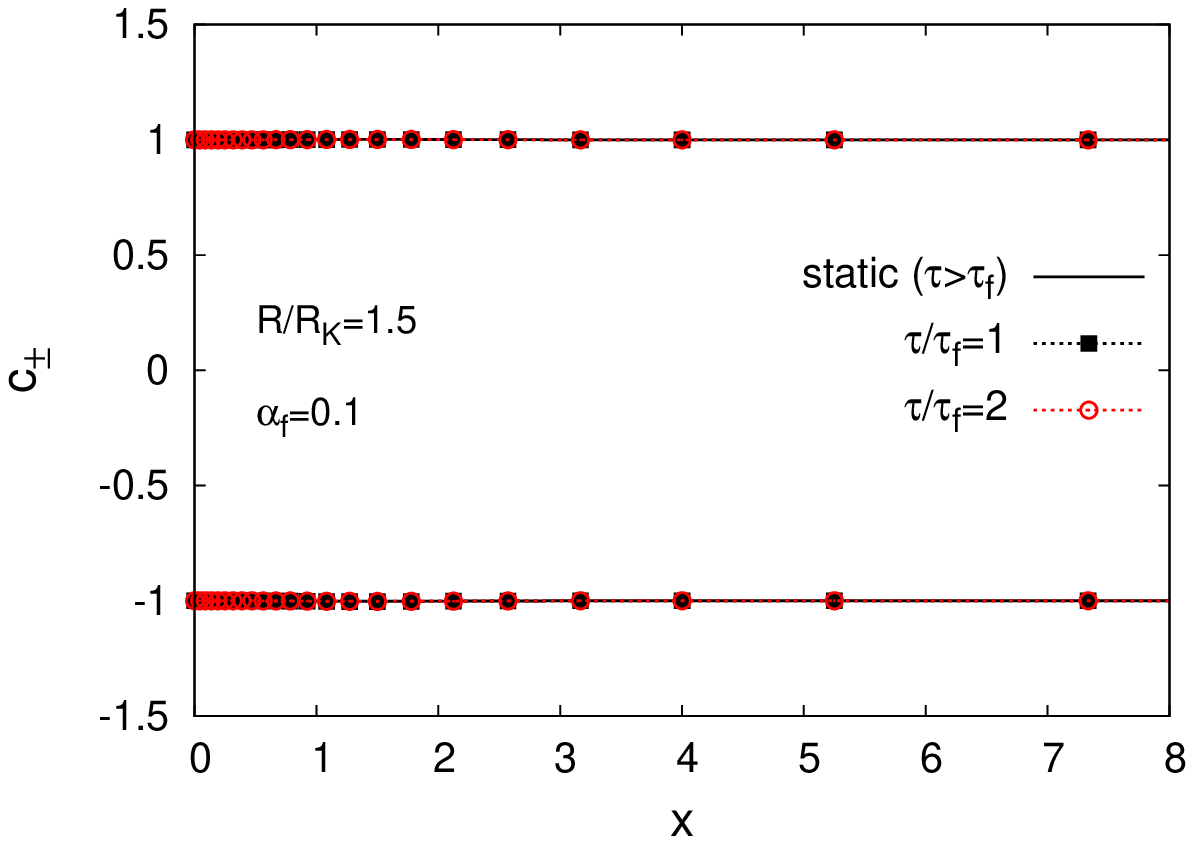}}
\end{center}
\caption{Time evolution of the scalar field derivatives and of the characteristic speeds,
for the polynomial kinetic function (\ref{K-K0-1-m-3}) and the Gaussian matter profiles
(\ref{eta-Gauss}), with $R/R_K=0.5$ ({\it left panels}) and $R/R_K=1.5$ ({\it right panels}),
starting with the initial condition $u=v=0$ at $\tau=0$. We choose a formation time-scale
$\tau_{\rm f}$ in Eq.(\ref{tauf-def}) with $\alpha_{\rm f}=0.1$.
{\it Upper panels:} time derivative $u(x,\tau)=\pl\phi/\pl\tau$ as a function of radius,
at times $\tau=0.25, 0.5, 1$ and $2\times \tau_{\rm f}$.
{\it Middle panels:} spatial derivative $v(x,\tau)=\pl\phi/\pl x$.
The solid lines are the quasistatic profiles defined by Eq.(\ref{static-u-v-x}) at $\tau=0.25, 0.5$ and
$1\times \tau_{\rm f}$ [these quasistatic solutions remain equal to the final static solution
defined by Eqs.(\ref{eta-Gauss}) and (\ref{static-u-v-x}) after $\tau_{\rm f}$].
{\it Lower panels:} characteristic speeds $c_{\pm}(x,\tau)$ from Eq.(\ref{cpm-def}).
The solid lines are the results (\ref{cpm-static}) on the final static profile (\ref{eta-Gauss}).}
\label{fig_Km3-G-t--uvc}
\end{figure*}

As an example of a nonstandard kinetic function $K(\chi)$ such that $W_{-}(y)$ is monotonically increasing up to $+\infty$, so that there is a unique well-defined static solution for any matter
density profile, which is also linearly stable to radial fluctuations, we consider the model
(\ref{K-power-1}) with $K_0=1$ and $m=3$,
\beq
K(\chi) = -1 + \chi + \chi^3 .
\label{K-K0-1-m-3}
\eeq

In Fig.~\ref{fig_Km3-G-t--uvc} we plot the evolution with time of the fields $u(x,\tau)$,
$v(x,\tau)$, and $c_{\pm}(x,\tau)$, for the Gaussian matter profile (\ref{eta-Gauss}),
with either $R/R_K=0.5$ or $R/R_K=1.5$.
This matter overdensity is built up over the time $\tau_{\rm f}$, starting at $\tau=0$, as in
Eq.(\ref{eta-tau}), and we choose $\alpha_{\rm f}=0.1$ in Eq.(\ref{tauf-def}).
We also take for the initial condition of the scalar field $u=v=0$ at $\tau=0$.

In both cases, $R/R_K=0.5$ or $R/R_K=1.5$, we find that the scalar field relaxes to the static
solution (\ref{static-u-v-x}) at late times, $\tau > \tau_{\rm f}$.
At earlier times, the scalar field approximately follows the evolution of the matter density profile
through the sequence of quasistatic states (\ref{static-u-v-x}).
There is a small but noticeable mismatch because the time-scale $\tau_{\rm f}$ is rather short
($\alpha_{\rm f}=0.1$ is not an extremely small number), so that the scalar field does not have the
time to fully relax to the quasistatic solutions until about $2 \tau_{\rm f}$.
This incomplete relaxation at early times mainly appears in the scalar field time-derivative $u$
and the characteristic speeds $c_{\pm}$, whereas the spatial gradient $v$ is always rather close to
the quasistatic solution.

At large distance, the matter density and the scalar field vanish and we recover the weak-field
limit, with characteristic speeds $c_{\pm}=\pm 1$.

For small values of $R/R_K$ (left panels), associated with high overdensities $\eta$ and
the nonlinear screened regime, we probe the nonlinearities of the kinetic  function $K$.
Then, the gradients of the scalar field are large (of order unity or
greater) and the characteristic speeds $c_{\pm}$ significantly depart from $\pm 1$.
For large values of $R/R_K$ (right panels), associated with low overdensities $\eta$ and
the linear unscreened regime, we mostly probe the linear part of the kinetic  function $K$
and nonlinearities are very weak.
Then, the gradients of the scalar field are smaller than unity
and the characteristic speeds $c_{\pm}$ always remain close to $\pm 1$.

\subsubsection{Relaxation for the DBI$^{+}$ model: $W(y)$ monotonically increases to $+\infty$
over $0 \leq y < y_-$ with a finite $y_-$}
\label{sec:relax-DBI+}

\begin{figure*}
\begin{center}
\epsfxsize=8.5 cm \epsfysize=5.8 cm {\epsfbox{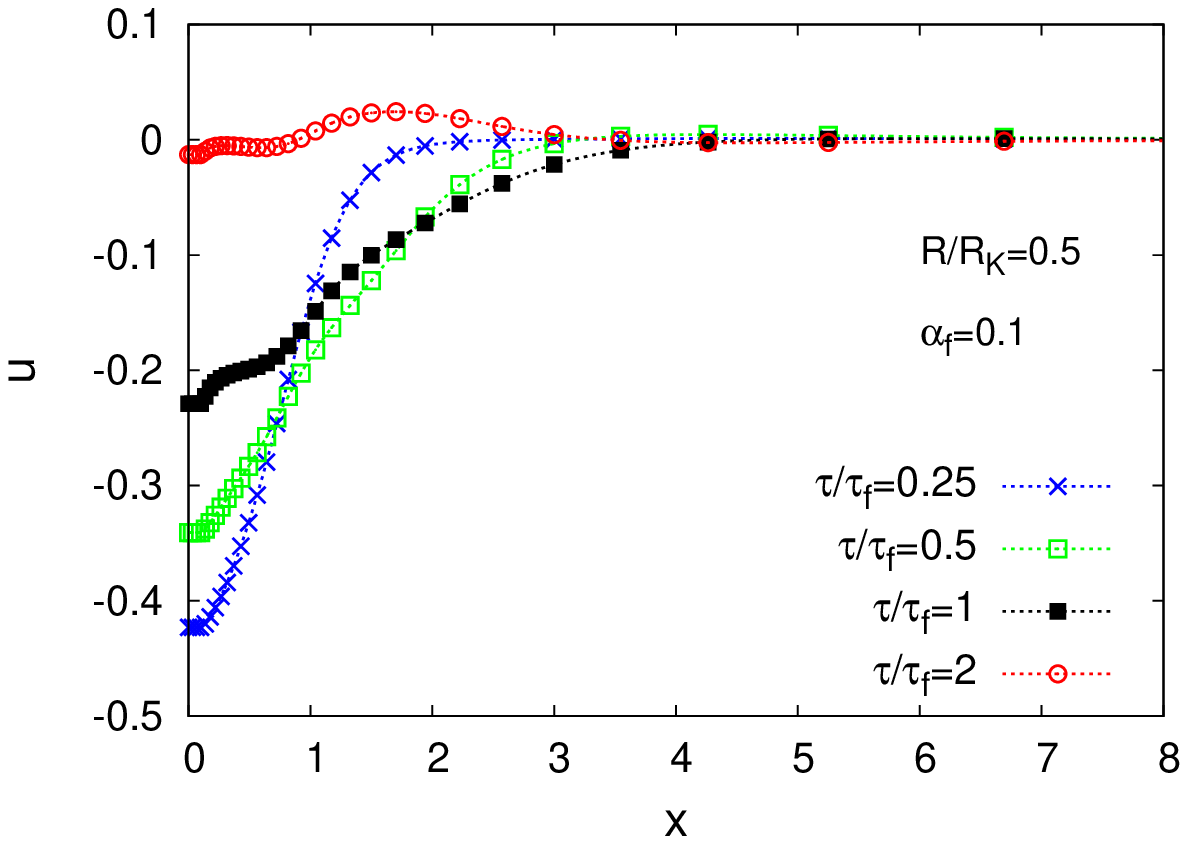}}
\epsfxsize=8.5 cm \epsfysize=5.8 cm {\epsfbox{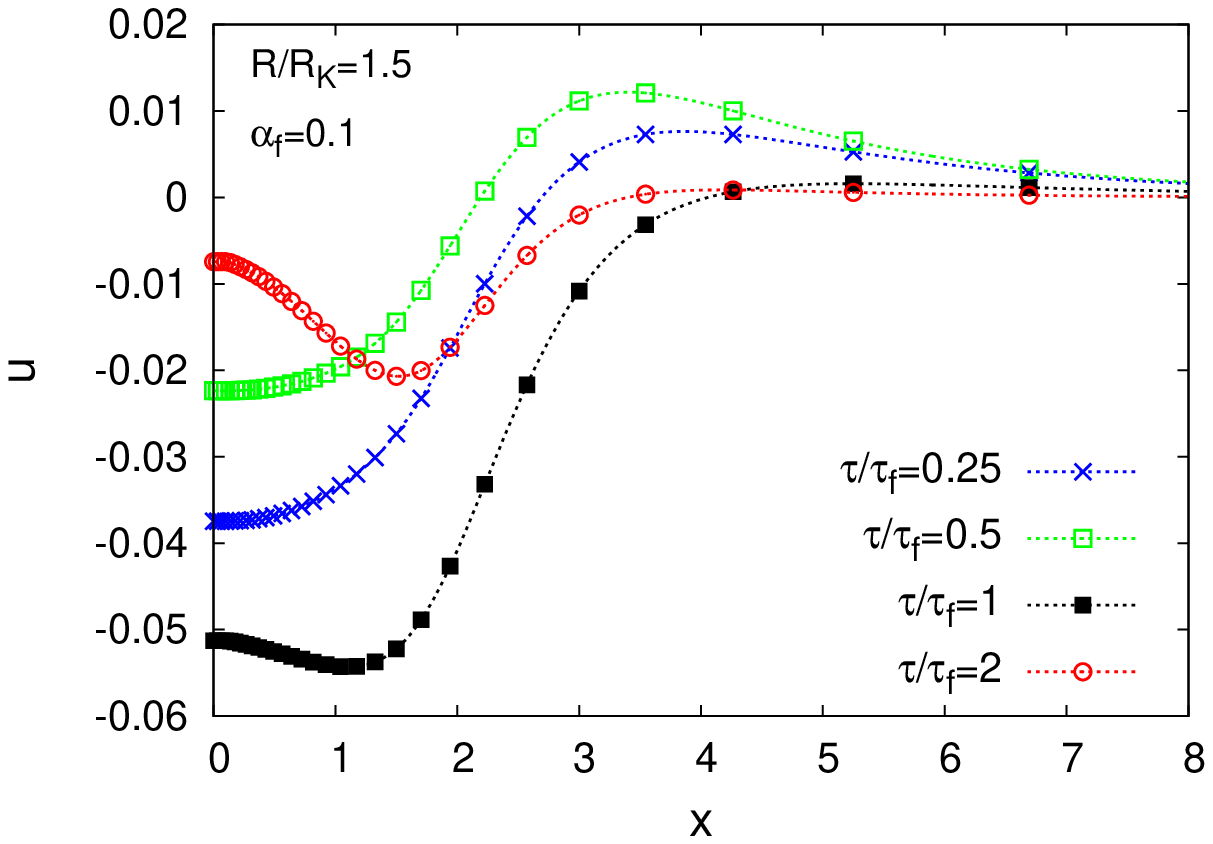}}\\
\epsfxsize=8.5 cm \epsfysize=5.8 cm {\epsfbox{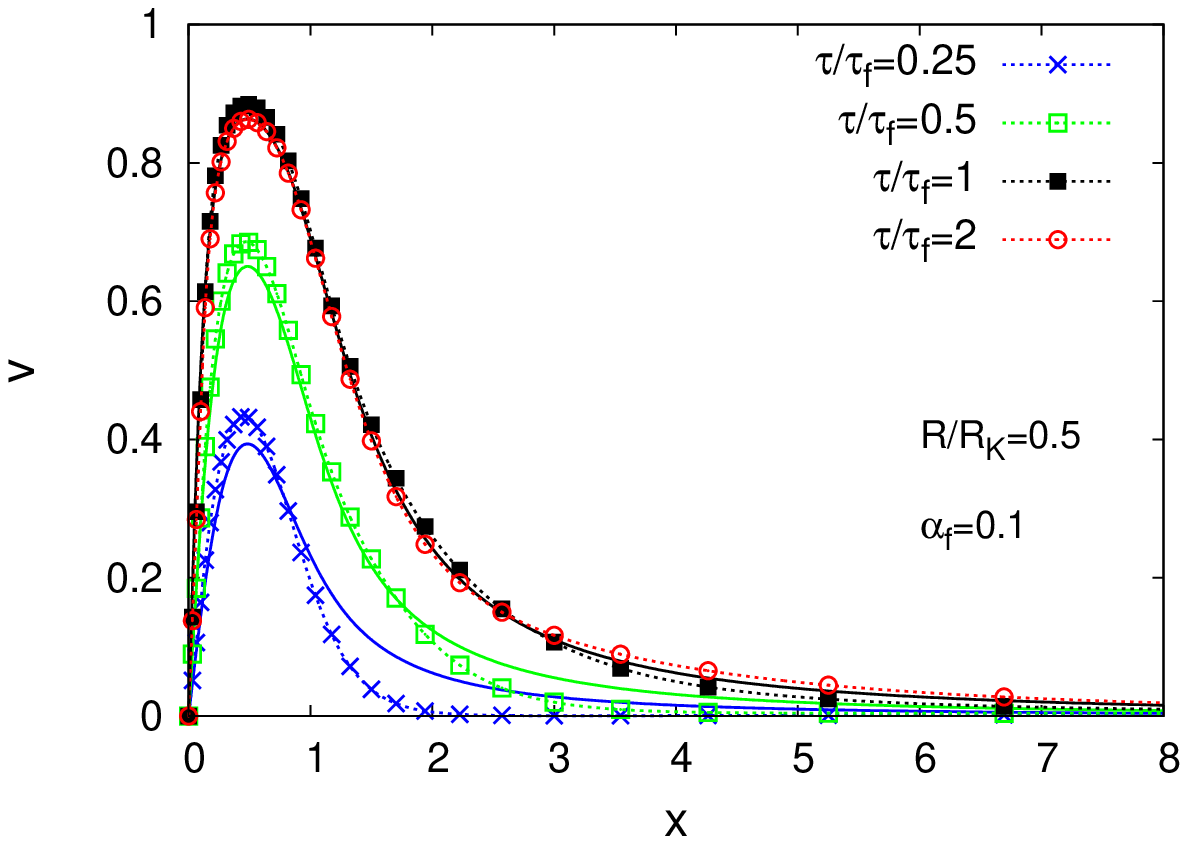}}
\epsfxsize=8.5 cm \epsfysize=5.8 cm {\epsfbox{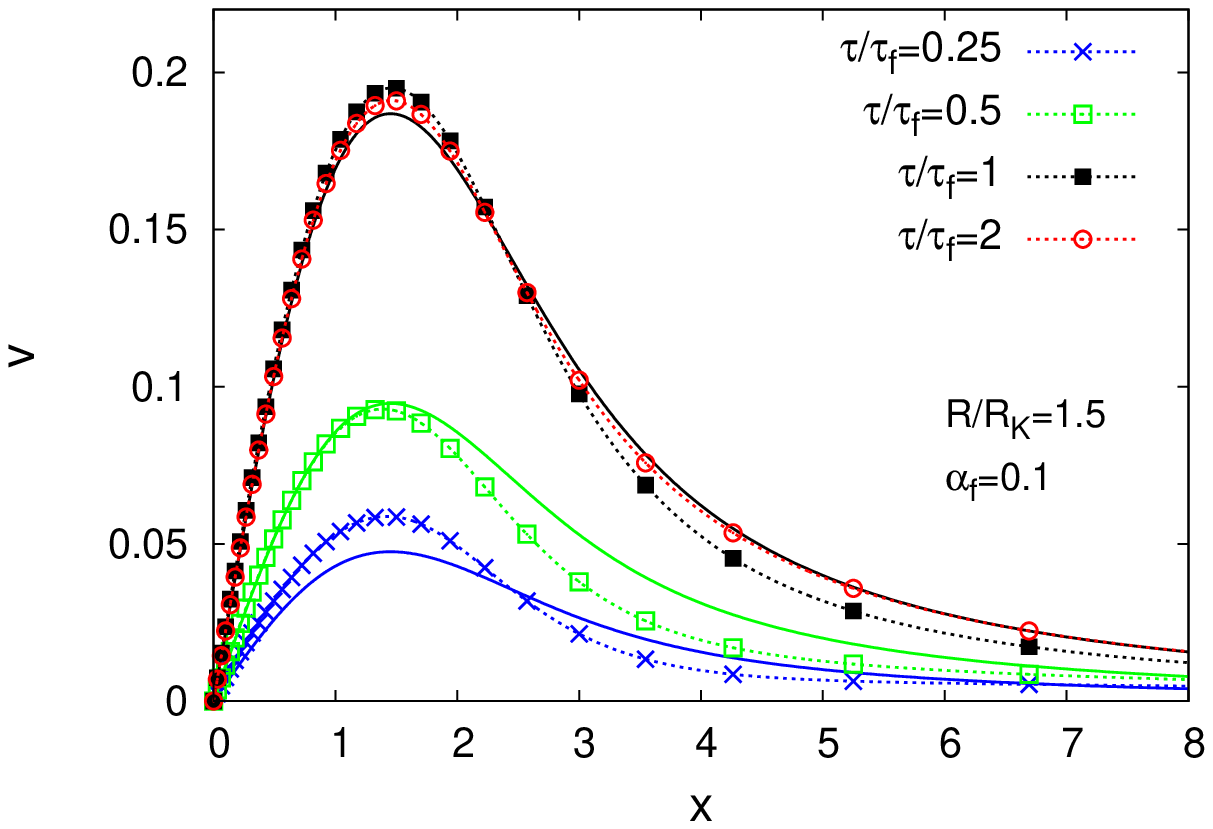}}\\
\epsfxsize=8.5 cm \epsfysize=5.8 cm {\epsfbox{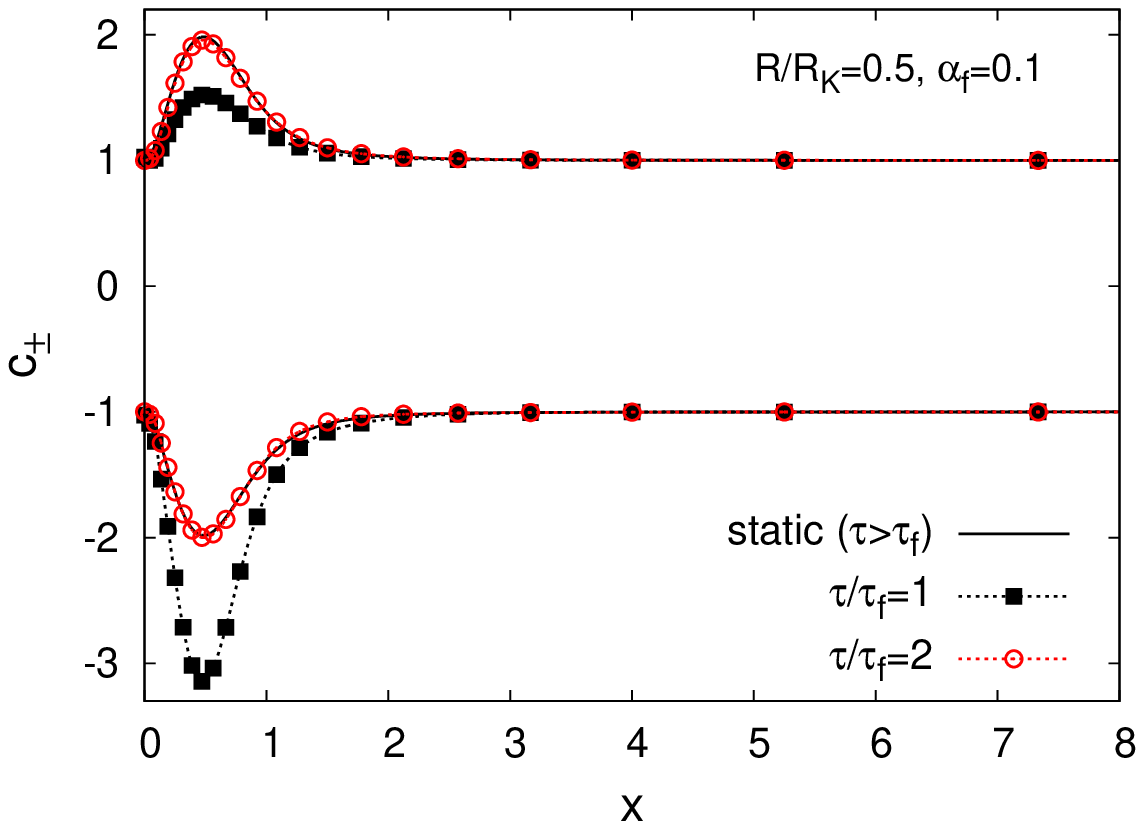}}
\epsfxsize=8.5 cm \epsfysize=5.8 cm {\epsfbox{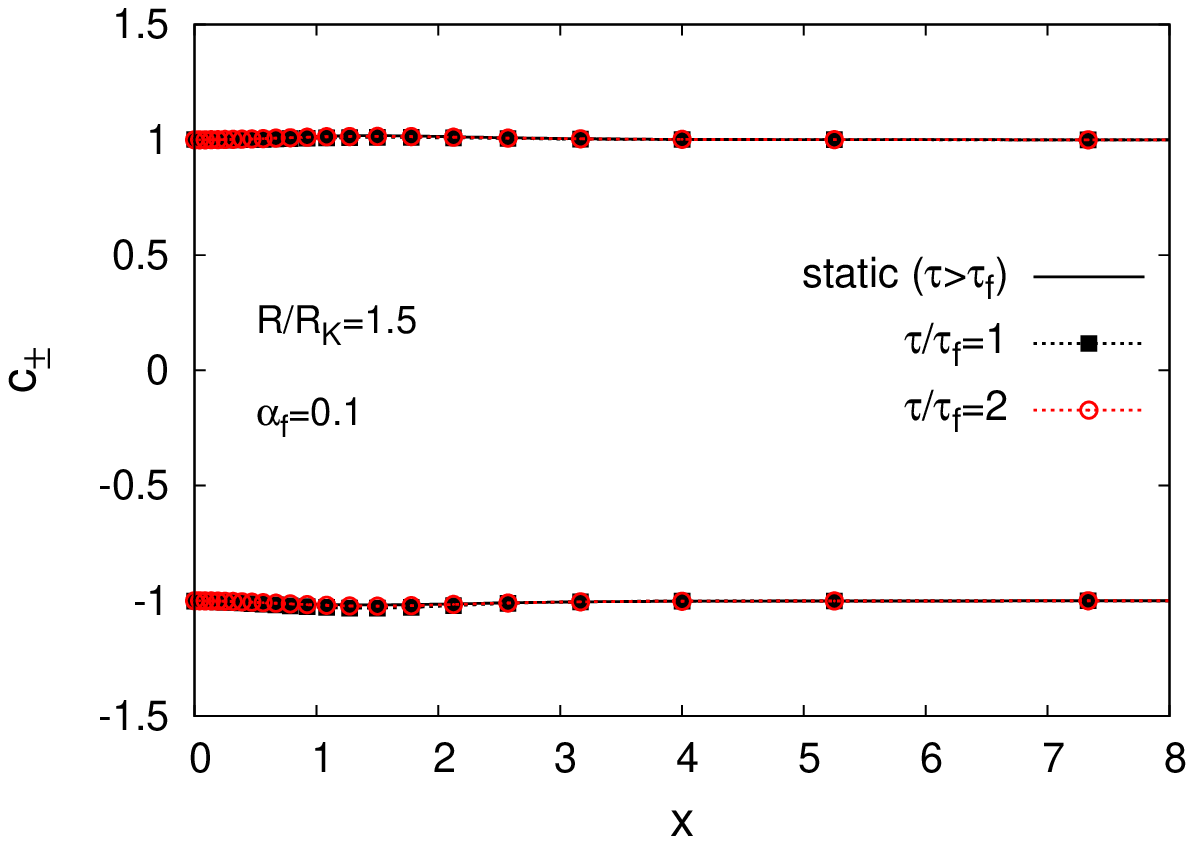}}

\end{center}
\caption{Time evolution of the scalar field derivatives and of the characteristic speeds,
for the ``DBI$^{+}$'' kinetic function (\ref{K-DBI+}) and the Gaussian matter profiles
(\ref{eta-Gauss}), with $R/R_K=0.5$ ({\it left panels}) and $R/R_K=1.5$ ({\it right panels}),
starting with the initial condition $u=v=0$ at $\tau=0$ and with $\alpha_{\rm f}=0.1$.
{\it Upper panels:} time derivative $u(x,\tau)=\pl\phi/\pl\tau$ as a function of radius,
at times $\tau=0.25, 0.5, 1$ and $2\times\tau_{\rm f}$.
{\it Middle panels:} spatial derivative $v(x,\tau)=\pl\phi/\pl x$.
The solid lines are the quasistatic profiles defined by Eq.(\ref{static-u-v-x}).
{\it Lower panels:} characteristic speeds $c_{\pm}(x,\tau)$ from Eq.(\ref{cpm-def}).
The solid lines are the results (\ref{cpm-static}) on the final static profile.}
\label{fig_KDBIp-G-t--uvc}
\end{figure*}

We now consider the case of the ``DBI$^{+}$'' model (\ref{K-DBI+def})
\cite{Burrage2014}, where
\beq
\mbox{DBI}^{+} \; : \;\;\; K(\chi)=\sqrt{1+2\chi}-2 , \;\;\; W_{-}(y) = \frac{y}{\sqrt{1-y^2}} .
\label{K-DBI+}
\eeq
This is an example of the cases where the function $W_{-}(y)$ is again monotonically increasing
up to $+\infty$, hence there is a well-defined static profile, but $y_-$ is finite (here $y_-=1$),
which implies that the scalar field spatial gradients are bounded.
We again consider the Gaussian matter profiles (\ref{eta-Gauss}), with $R/R_K=0.5$ and
$1.5$, in Fig.~\ref{fig_KDBIp-G-t--uvc}.

Because of the divergence of $W_{-}(y)$ at the upper bound $y_-=1$ the nonlinearities are
stronger than in the polynomial case studied in Sec.~\ref{sec:relax-monotonic}
and appear at greater values of $R/R_K$ (as shown by the comparison of the characteristic
speeds $c_{\pm}$ in the case $R/R_K=0.5$).
Nevertheless, we obtain the same qualitative behavior as for the polynomial kinetic function
(\ref{K-K0-1-m-3}). In both cases, the scalar field relaxes to the static profile (\ref{static-u-v-x}).
For large values of $R/R_K$, in the unscreened linear weak-field regime, the characteristic speeds
$c_{\pm}$ always remain close to $\pm 1$.
For small values of $R/R_K$, in the screened nonlinear strong-field regime, the characteristic speeds
significantly depart from $\pm 1$.
However, because $\alpha_{\rm f}\ll 1$ the scalar field is able to follow the evolution of the matter
profile by approximately going through the sequence of quasistatic solutions and no shocks appear.

In contrast, our numerical computations of the extreme case $\tau_{\rm f}=0$, where the evolution
is far from quasistatic, again show that for small values of $R/R_K$, in the screened nonlinear
strong-field regime, the time evolution is much more violent. Then, transient shocks appear
with characteristic speeds that significantly depart from $\pm 1$ and show non-monotonic behaviors.
Moreover, the relaxation does not proceed in a uniform manner
(at small radii the field gradients significantly ``overshoot'' the static gradients  whereas at
large radii they smoothly converge to the static value).
However, we find that at late times the scalar field again relaxes to the final static solution.

\subsubsection{Relaxation and runaway for the DBI$^{-}$ model: $W_{-}(y)$ is bounded}
\label{sec:relax-DBI-}

\begin{figure*}
\begin{center}
\epsfxsize=8.5 cm \epsfysize=5.8 cm {\epsfbox{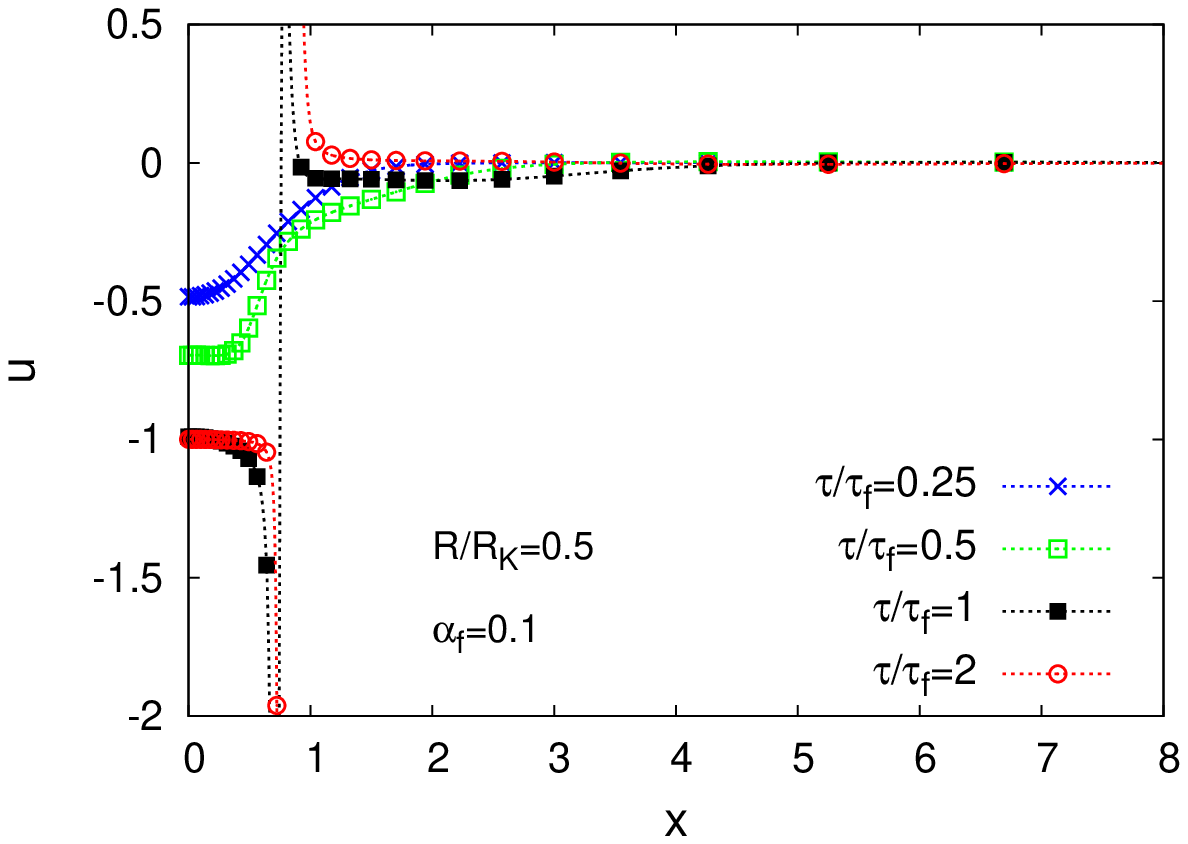}}
\epsfxsize=8.5 cm \epsfysize=5.8 cm {\epsfbox{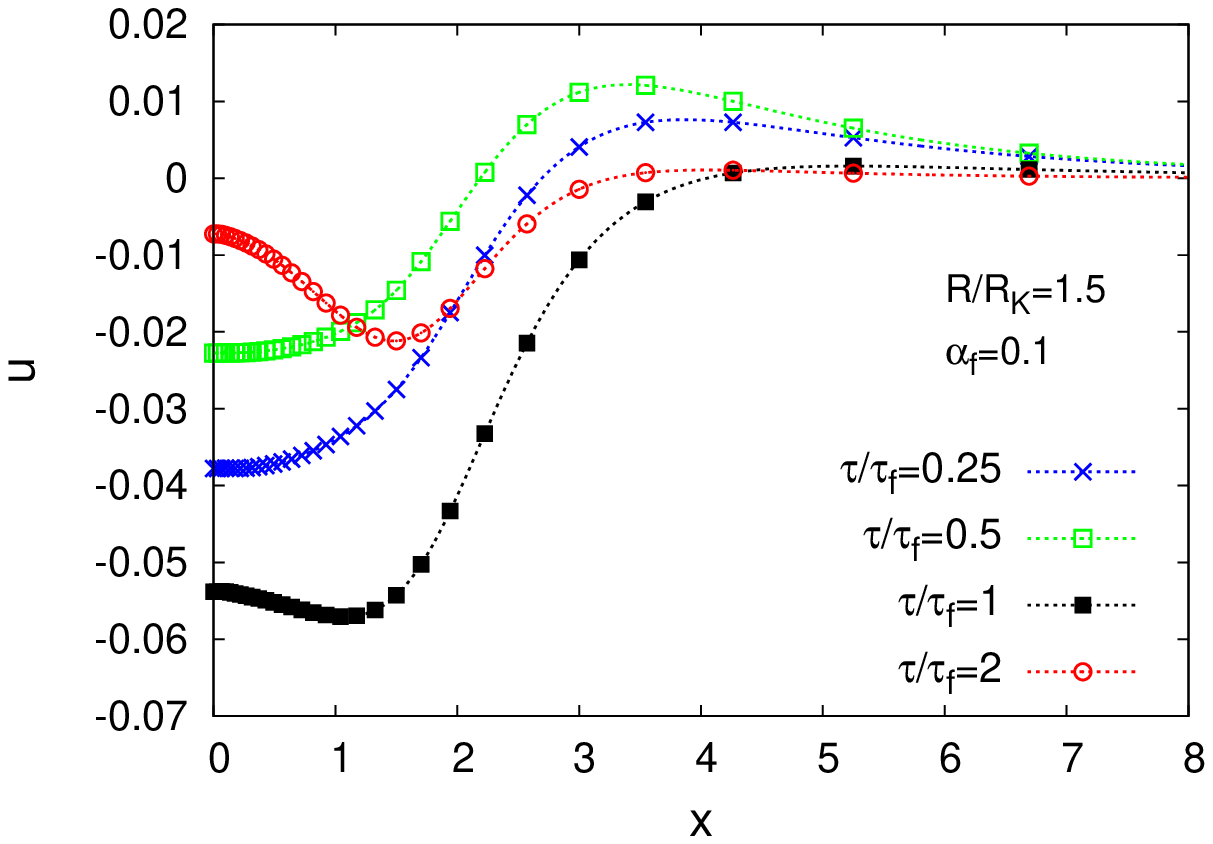}}\\
\epsfxsize=8.5 cm \epsfysize=5.8 cm {\epsfbox{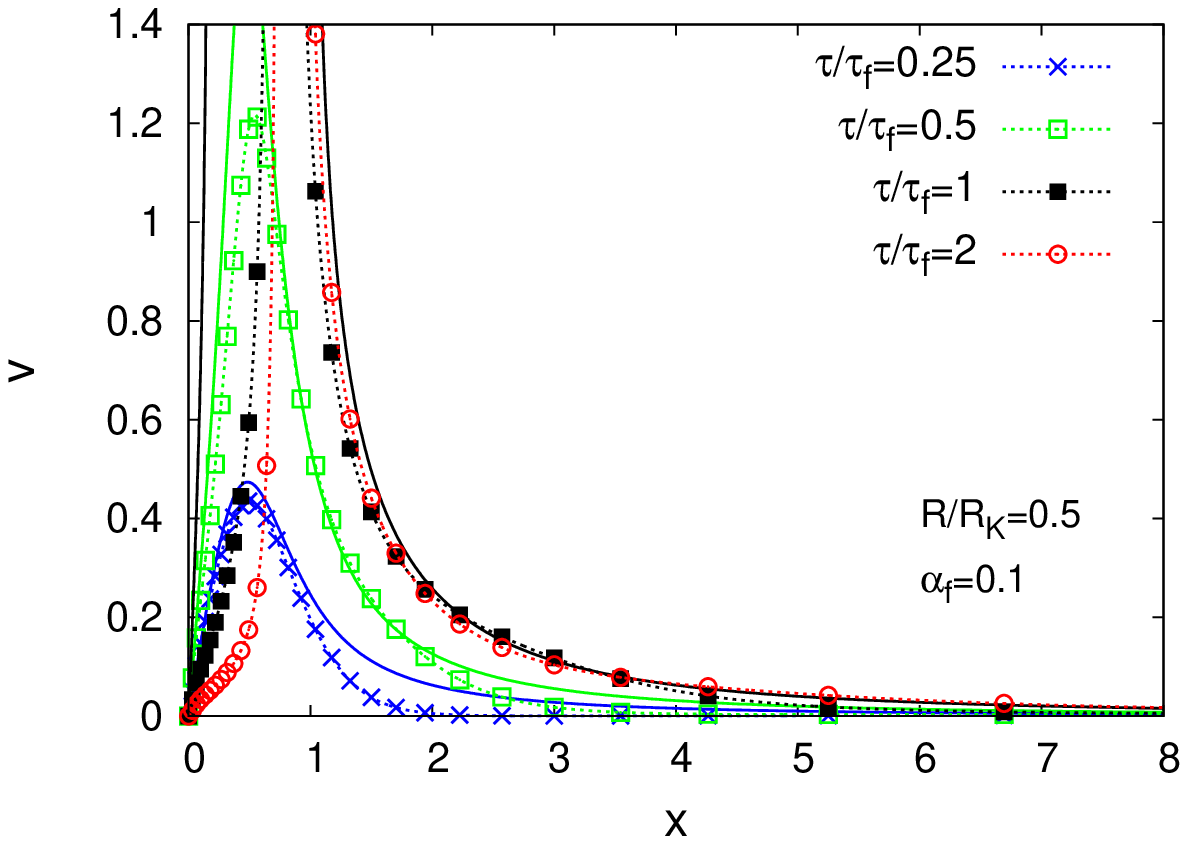}}
\epsfxsize=8.5 cm \epsfysize=5.8 cm {\epsfbox{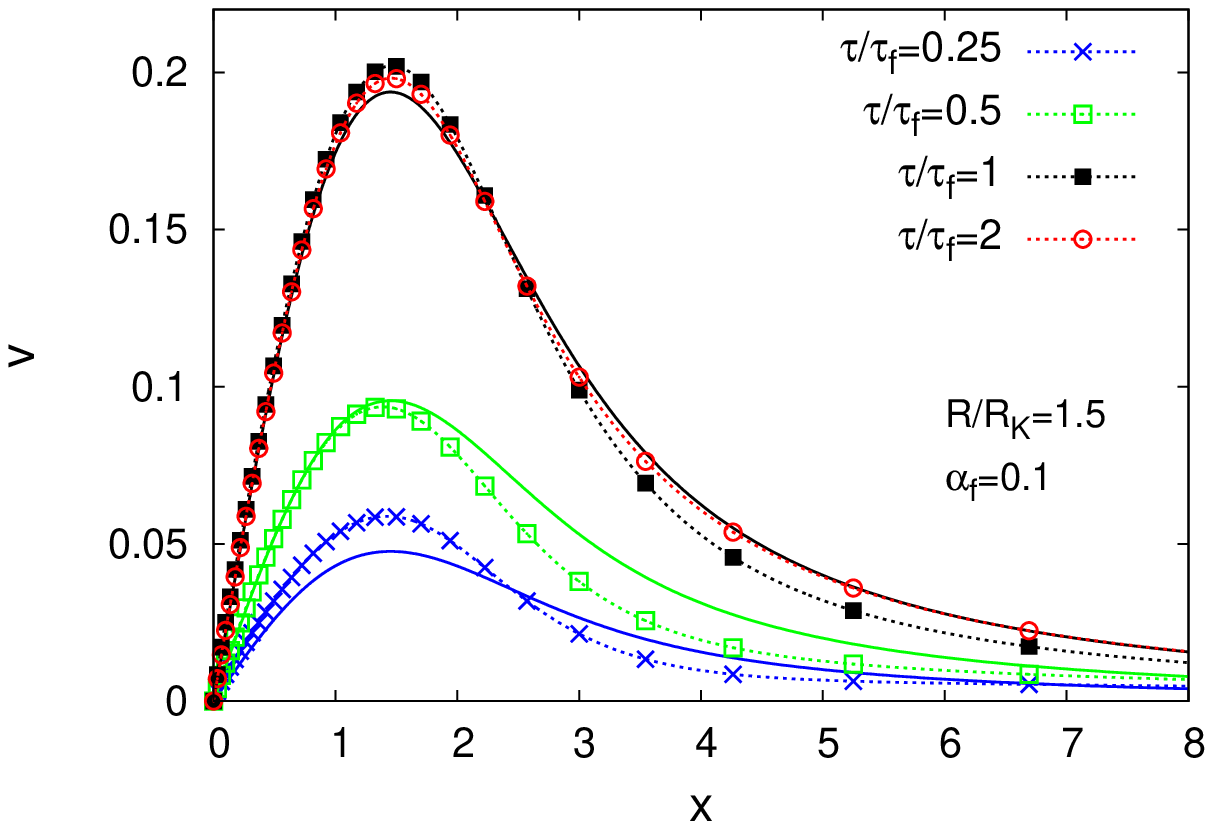}}\\
\epsfxsize=8.5 cm \epsfysize=5.8 cm {\epsfbox{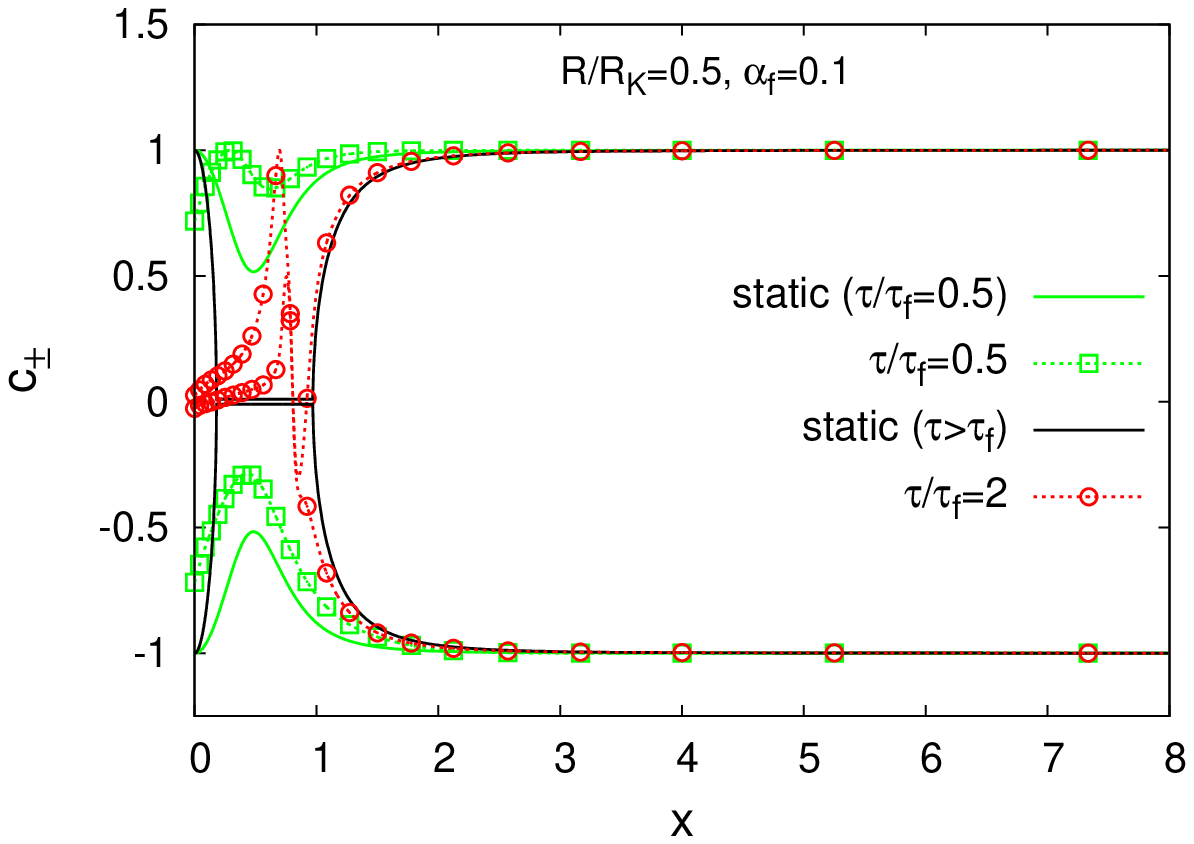}}
\epsfxsize=8.5 cm \epsfysize=5.8 cm {\epsfbox{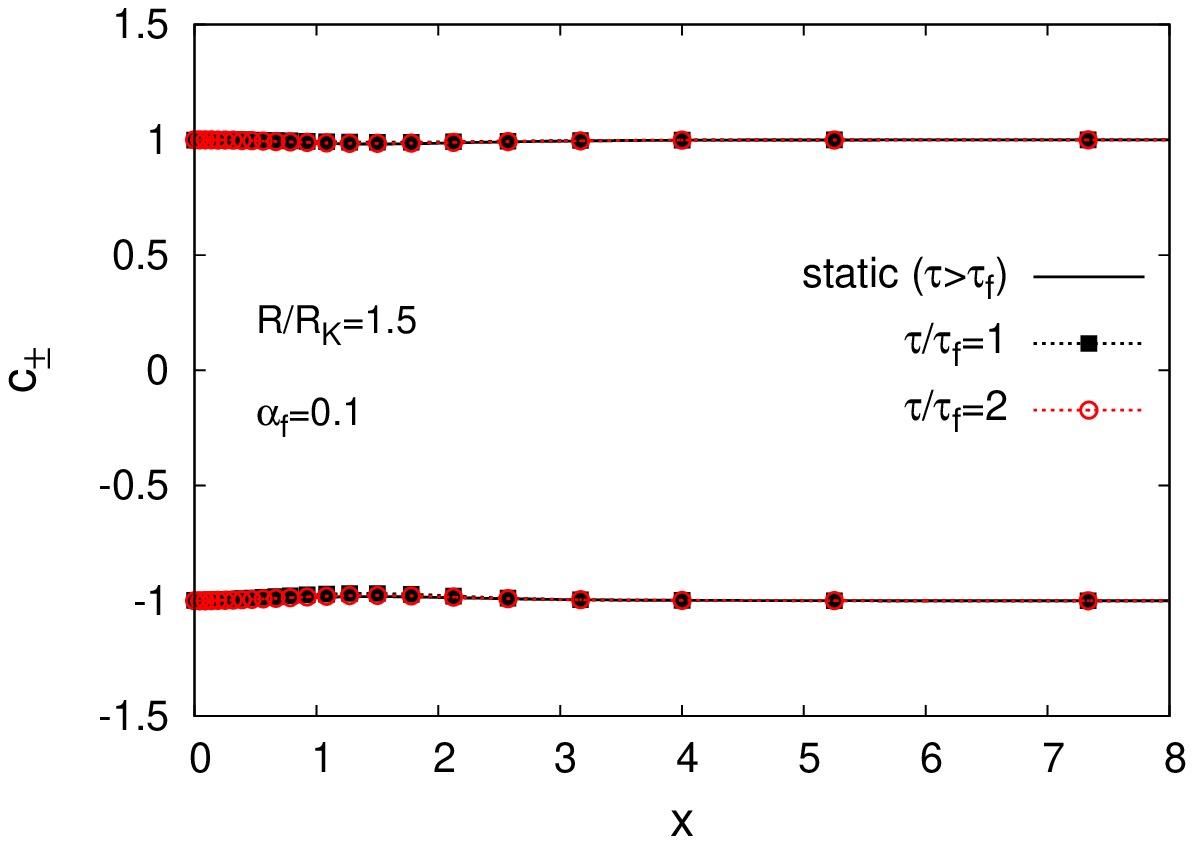}}

\end{center}
\caption{Time evolution of the scalar field derivatives and of the characteristic speeds,
for the ``DBI$^{-}$'' kinetic function (\ref{K-DBI-}) and the Gaussian matter profiles (\ref{eta-Gauss}),
with $R/R_K=0.5$ ({\it left panels}) and $R/R_K=1.5$ ({\it right panels}),
starting with the initial condition $u=v=0$ at $\tau=0$ and with $\alpha_{\rm f}=0.1$.
{\it Upper panels:} time derivative $u(x,\tau)=\pl\phi/\pl\tau$ as a function of radius,
at times $\tau=0.25, 0.5, 1$ and $2\times\tau_{\rm f}$.
{\it Middle panels:} spatial derivative $v(x,\tau)=\pl\phi/\pl x$.
The solid lines are the quasistatic profiles defined by Eq.(\ref{static-u-v-x}).
{\it Lower panels:} characteristic speeds $c_{\pm}(x,\tau)$ from Eq.(\ref{cpm-def}).
The solid lines are the results (\ref{cpm-static}) on the final static profile.}
\label{fig_KDBIm-G-t--uvc}
\end{figure*}

Finally, we consider the case of the ``DBI$^{-}$'' model (\ref{K-DBI-def}), where
\beq
\mbox{DBI}^{-} \; : \;\;\; K(\chi)= -\sqrt{1-2\chi} , \;\;\; W_{-}(y) = \frac{y}{\sqrt{1+y^2}} .
\label{K-DBI-}
\eeq
This is an example of the cases where the function $W_{-}(y)$ is monotonically increasing
but shows a finite upper bound.
As seen from the analysis in Secs.~\ref{sec:static-spherical} and \ref{sec:bounded},
this means that for low overdensities,
where we probe the linear regime of the kinetic function, we find a well-defined static profile,
but for high overdensities, where the right-hand side in Eq.(\ref{KG-W-1}) can reach values that
are greater than the maximum of $W_{-}(y)$, there is no static solution that is valid throughout
all space.
We again consider the Gaussian matter profiles (\ref{eta-Gauss}), with $R/R_K=0.5$ and
$1.5$, and we display the time evolution in Fig.~\ref{fig_KDBIm-G-t--uvc}.

For large values of $R/R_K$ (right panels), in the weak-field regime where we do not probe the
upper bound of $W_{-}(y)$, we obtain the same qualitative behavior as for
the polynomial kinetic function (\ref{K-K0-1-m-3}).
In both cases the scalar field relaxes to the static profile (\ref{static-u-v-x}) through a regular
time evolution and the characteristic speeds $c_{\pm}$ always remain close to $\pm 1$.

For small values of $R/R_K$, in the screened nonlinear strong-field regime, we are sensitive to
the upper bound of $W_{-}(y)$. Then, on intermediate scales $r \sim R$, where the ratio $m(x)/x^2$
(which is also given by the Newtonian force) is the greatest, there is no solution to the static
equation (\ref{KG-W-1}) at late times, $\tau \gtrsim 0.7 \times \tau_{\rm f}$, when the matter overdensity
has reached large-enough values.
Thus, for $\tau \geq \tau_{\rm f}$ the static solution $v_{\rm static}(x)$ shown in the middle
left panel is only defined over the two intervals $[0,x_-[$ and $]x_+,+\infty[$, with
$v(x) = y(x) \rightarrow\infty$ for $x\rightarrow x_{\pm}$,
whereas for $\tau=0.25$ and $0.5\times\tau_{\rm f}$ the quasistatic solutions are well defined over
all space.
Moreover, from Eq.(\ref{cpm-static}), on the quasistatic profiles the characteristic speeds are
$c_{\pm}=\pm 1/\sqrt{1+y^2}$. Therefore, for $\tau \gtrsim 0.7 \times \tau_{\rm f}$ the characteristic
speeds vanish at the boundaries of the ill-defined region,
$c_{\pm}(x) \rightarrow 0$ for $x\rightarrow x_{\pm}$.

Then, we can see on the left panels in Fig.~\ref{fig_KDBIm-G-t--uvc} that at early times,
$\tau \lesssim 0.7 \tau_{\rm f}$, when the quasistatic solutions are well defined over all space,
the scalar field approximately follows these quasistatic solutions as in the right panels and
as in the cases displayed in Figs.\ref{fig_Km3-G-t--uvc} and \ref{fig_KDBIp-G-t--uvc}.
At later times, when the quasistatic solutions are only defined on two disjoint intervals
$[0,x_-[$ and $]x_+,+\infty[$, the scalar field keeps following the quasistatic solution on the
outer range $]x_+,+\infty[$ but deviates from it on the inner range $[0,x_-[$ while shocks
appear in the ill-defined region $]x_-,x_+[$.
Then, at late times $\tau>\tau_{\rm f}$, when the matter profile no longer evolves,
the scalar field relaxes to the static profile on the outer interval $]x_+,+\infty[$, while it never
reaches a static state at smaller radii $x<x_+$, including in the inner region $[0,x_-[$ where a
static profile could be defined. As could be expected, the scalar field spatial gradient
$v(x,\tau)$ keeps increasing at radii $x \lesssim x_+$, in the endless attempt to find
large-enough values of $W_{-}$ to satisfy Eq.(\ref{KG-W-1}), and the time derivative $u(x,\tau)$
does not converge to zero.
At small radii we have $u(x) \simeq -1$ because the spatial derivative $v(x)$ vanishes
for $x\rightarrow 0$ and $\chi=(u^2-v^2)/2$ is bounded from above by $1/2$,
as seen from the definition (\ref{K-DBI-}) of the DBI$^{-}$ kinetic function.
Again, the nonlinearities lead to the formation of shocks, which remain present at late times
in the inner region below $x_+$ where there is no convergence to a static state.

This numerical analysis also shows that the system does not converge towards the generalized
solutions of the form (\ref{nu-def}), with a large-enough uniform value $\nu$ for the time-derivative
of the scalar field, to achieve a steady-state solution that applies at all radii.
This is actually a nice feature because such states would imply a loss of predictability of the model,
as the coefficient $\nu$ could not have been predicted from the static matter density profile alone
and would have shown some dependence on the history of the system.

The convergence towards the static solution in the outer range $x>x_+$ means that if
we restrict to outer radii this DBI$^{-}$ model provides a well-defined static limit
and is a predictive model. However, the model is not completely predictive in the inner range
$x<x_-$ because of the runaway behavior, which depends on the initial conditions
(e.g., the time of formation of the matter overdensity and the initial scalar field state).
Because the gradients are large in this regime the fifth force is typically suppressed by the
factor $1/K'$ but since there is a crossover from the outer spatial part where $\chi <0$
to the inner time-like part where $\chi>0$ there exists a thin shell where the fifth force would
be large and even divergent.

Alternatively, we can assume that for large gradients the expression (\ref{K-DBI-}) is no
longer correct and higher-order contributions come into play that enable the relaxation
towards a static solution.
The results shown in Fig.~\ref{fig_KDBIm-G-t--uvc} suggest that this would only regularize
the inner region $0 \leq x \leq x_+$ while keeping unaffected the outer region
$x>x_+$, where the scalar field would remain on the static solution given by the ``low-gradient''
approximation (\ref{K-DBI-}).
In this framework, our results suggest that there would be an efficient separation of scales,
so that on large scales far from overdensities, in the moderate nonlinear regime, the scalar
field can be described by the `low-gradient'' kinetic function (\ref{K-DBI-}), whereas close
to high-density objects one must either go beyond the quasistatic approximation
(at the expense of some loss of predictivity, because of the dependence on the details
of the initial conditions) or introduce a higher-order regularisation.

Another shortcoming of this model is that for large negative $\chi$ the derivative $K'$ 
goes to zero. Therefore, there is no efficient screening in the static nonlinear regime, and the
fifth force actually becomes large as compared with the Newtonian force. Thus, this model 
cannot provide a realistic scenario.

\subsubsection{Models with $K'<0$ or $W_{-}'<0$}
\label{sec:relax-K-W-}

We do not consider models where $K'$ or $W_{-}'$ can reach negative values.
Indeed, this yields complex characteristic speeds at some point during the time
evolution, see Eqs.(\ref{cpm-def}) and (\ref{cpm-static}), and the system is no longer
hyperbolic.
This means that the discontinuous solutions found in Sec.~\ref{sec:discontinuous}
are not physical and must be disregarded.

We consider such models of $K(\chi)$ badly behaved and unrealistic for physical
applications.

\subsubsection{Top-hat matter density profiles}
\label{sec:top-hat}

We also computed the relaxation of the scalar field when, instead of the Gaussian profiles
(\ref{rho-Gauss})-(\ref{eta-Gauss}), the matter density obeys a top-hat profile, given by
\beq
r < R : \;\; \rho(r) = \rho_0  , \;\;\;\; r>R : \;\; \rho(r) = 0 ,
\label{rho-top-hat}
\eeq
which reads in terms of the dimensionless variables $x$ and $\eta$ as
\beq
x < \frac{R}{R_K} : \;\; \eta(x) = 3 \left( \frac{R_K}{R} \right)^3 , \;\;\;\;
x > \frac{R}{R_K} : \;\; \eta(x) = 0 .
\label{eta-top-hat}
\eeq
We considered the same three kinetic functions
as in Secs.~\ref{sec:relax-monotonic}, \ref{sec:relax-DBI+}, and
\ref{sec:relax-DBI-}.

We always find similar behaviors to the cases of the Gaussian matter profiles (\ref{rho-Gauss}),
with a relaxation towards the static solution when it exists. When the static solution only exists
on two disjoint intervals $[0,x_-[$ and $]x_+,+\infty[$, we again find that the scalar field
converges to the static solution on the outer range. When the size of the object is
greater than its K-mouflage radius, $R \gtrsim R_K$, we probe the unscreened weak-field regime
and the characteristic speeds $c_{\pm}$ remain close to $\pm 1$.
When the size of the object is smaller than its K-mouflage radius, $R \lesssim R_K$, we probe the
screened strong-field regime and the characteristic speeds $c_{\pm}$ show significant departures
from $\pm 1$.

\subsubsection{Very large or small formation time for the matter density profile}
\label{sec:tauf-0}

We also computed the relaxation of the scalar field when the formation time-scale $\tau_{\rm f}$
of the matter profile is ten times greater than in
Figs.\ref{fig_Km3-G-t--uvc}, \ref{fig_KDBIp-G-t--uvc}, and \ref{fig_KDBIm-G-t--uvc},
that is, $\alpha_{\rm f}=0.01$.
As expected, we find that the scalar field follows the quasistatic solutions to a greater accuracy,
because it has more time to relax as compared with the evolution time-scale of the matter profile.
At late times, it again relaxes to the static solution when the latter exists over all space.
In the DBI$^-$ model of Fig.~\ref{fig_KDBIm-G-t--uvc}, when the final static solution is
only defined on two disjoint intervals $[0,x_-[$ and $]x_+,+\infty[$, we again find that the scalar field
only converges to the static solution on the outer range, while shocks appear at smaller radii and
gradients keeps growing with time.

We also considered the opposite case where the matter density profile (\ref{eta-Gauss})
is built instantaneously, i.e. $\tau_{\rm f}=0$.
This is a somewhat academic exercise, because in practice the formation time-scale
$\tau_{\rm f}$ is expected to be of the order or greater than the scalar-field relaxation time-scale
$\tau_{\phi}$, see Eq.(\ref{tauf-def}).
However, on small scales for non-spherical
configurations it might happen that the scalar field takes more time to relax than is suggested by
the simple dimensional analysis of Eq.(\ref{tauf-def}), therefore it remains interesting to check the
limiting case $\tau_{\rm f}=0$.

Then, our numerical computations show that in the nonlinear regime (i.e., for small values of $R/R_K$),
the relaxation proceeds in a more violent fashion than the quasistatic evolution associated with
the cases $\alpha_{\rm f} \ll 1$.
In particular, strong time and spatial gradients of the scalar field appear at transient stages
and the characteristic speeds $c_{\pm}$ show strong deviations from $\pm 1$ and non-monotonic
regions.
As usual for nonlinear transport equations, this gives rise to transient shocks at small radii,
in the nonlinear regime.
At late times, these transient shocks disappear as the scalar field relaxes to
the static solution.
For large values of $R/R_K$,
even with an instantaneous matter density structure formation, no shocks appear and the
relaxation towards the static solution proceeds in a very regular manner, as we only probe
the linear unscreened regime and the linear part of the kinetic  function $K$ and
the characteristic speeds always remain close to $\pm 1$ (which prevents shock formation).

Again, in the DBI$^-$ model of Fig.~\ref{fig_KDBIm-G-t--uvc}, when the final static solution is
only defined on two disjoint intervals $[0,x_-[$ and $]x_+,+\infty[$, the scalar field
only converges to the static solution on the outer range, while at smaller radii shocks are present
at all times and gradients keeps growing with time.

\section{A Summary of K-mouflage Properties}
\label{sec:Summary}

We summarize below the main properties of K-mouflage models that we have obtained,
both for the background cosmology studied in \cite{Brax:2014aa,Brax:2014ab}
and for the small-scale screening regime described in this paper.
We first describe power-law and DBI-like models, and next give the general results that we
have obtained.

\subsection{Cosmological properties along $\chi>0$}
\label{sec:Cosmological}

\subsubsection{$K(\chi) \sim K_0 \chi^m$ with $K_0>0$ and $m>1$}

The constraint $m>1$ arises from the requirement that $\bar{\rho}_{\varphi} \ll \bar\rho$
at early times (i.e., we recover the matter-dominated Einstein-de Sitter expansion).

\begin{itemize}

\item  At early times, the background field satisfies $\bar\varphi<0$ and $\dot{\bar\varphi}<0$,
$\bar\rho_{\varphi} > 0$, $\bar\rho_{\varphi}^{\rm eff} < 0$.
More precisely, at $t \ll t_0$ far in the matter-dominated era, we have
$\dot{\bar\varphi} \sim - t^{-1/(2m-1)}$,
$\bar\rho_{\varphi} \sim - \bar\rho_{\varphi}^{\rm eff} \sim t^{-2m/(2m-1)}$.

\item The effective equation of state parameter has $w_{\varphi}^{\rm eff} < -1$ at low redshift.

\item  There are no ghosts (because $\bar{K}'>0$ and $\bar{K}'+2\bar{\chi}\bar{K}''>0$).

\item The formation of large-scale cosmological structures is enhanced.

\end{itemize}

\subsubsection{$K(\chi) \sim K_0 \chi^m$ with $K_0<0$ and $m>1$}

\begin{itemize}

\item At early times, the background field satisfies $\bar\varphi>0$ and $\dot{\bar\varphi}>0$,
$\bar\rho_{\varphi} < 0$, $\bar\rho_{\varphi}^{\rm eff} > 0$.
More precisely, at $t \ll t_0$ we have $\dot{\bar\varphi} \sim t^{-1/(2m-1)}$,
$-\bar\rho_{\varphi} \sim \bar\rho_{\varphi}^{\rm eff} \sim t^{-2m/(2m-1)}$.

\item The effective equation of state parameter has $w_{\varphi}^{\rm eff} > -1$ at low redshift.

\item There are ghosts, which makes the model very contrived.

\item The formation of large-scale cosmological structures is suppressed.

\end{itemize}

\subsubsection{DBI$^{+}$ model}

From $K(\chi)= \sqrt{1+2\chi}-2$ we find that the background Klein-Gordon equation
can be integrated as
\beq
\frac{\dot{\bar\varphi}/{\cal M}^2}{\sqrt{1+\dot{\bar\varphi}^2/{\cal M}^4}}
= - \frac{\beta\bar\rho t}{{\cal M}^2 \MPl} .
\label{DBIp-cosmo}
\eeq
Because the left-hand side is bounded, this model cannot apply to high redshifts, where
the background matter density grows as $\bar\rho \sim t^{-2}$.
Therefore, one must either disregard this model or include higher-order corrections to the
kinetic function $K(\chi)$.
This shortcoming is related to the constraint $m>1$ found for generic power-law behaviors
$K(\chi) \sim \chi^m$ at $\chi\rightarrow +\infty$.

\subsubsection{DBI$^{-}$ model}

From $K(\chi)= -\sqrt{1-2\chi}$ we find that the background Klein-Gordon equation
can be integrated as
\beq
\frac{\dot{\bar\varphi}/{\cal M}^2}{\sqrt{1-\dot{\bar\varphi}^2/{\cal M}^4}}
= - \frac{\beta\bar\rho t}{{\cal M}^2 \MPl} .
\label{DBIm-cosmo}
\eeq
The left-hand side goes to $-\infty$ for $\dot{\bar\varphi} \rightarrow - {\cal M}^2$, hence
this model can be extended up to any redshift.

\begin{itemize}

\item  At early times, the background field satisfies $\bar\varphi<0$, $\dot{\bar\varphi}<0$,
and $\bar\rho_{\varphi} > 0$.
More precisely, at $t \ll t_0$ we have $\dot{\bar\varphi} \simeq - {\cal M}^2$,
$\bar\rho_{\varphi} \sim t^{-1}$, and$|\bar\rho_{\varphi}^{\rm eff}| \ll t^{-1}$.

\item The effective equation of state parameter has $w_{\varphi}^{\rm eff} < -1$ at low redshift.

\item  There are no ghosts (because $\bar{K}'>0$ and $\bar{K}'+2\bar{\chi}\bar{K}''>0$).

\item The formation of large-scale cosmological structures is enhanced.

\end{itemize}

\subsection{Small-scale properties along $\chi <0$}

\subsubsection{$K(\chi) \sim K_0 \chi^m$ with $K_0>0$ and $m$ odd, or $K_0<0$ and $m$ even, so that $K'>0$}

\begin{itemize}

\item The function $W_{-}(y)$ is monotonically increasing up to $+\infty$ over $0 \leq y <+\infty$
and there is a unique well-defined static scalar field profile for any static matter density profile.

\item The fifth force amplifies Newtonian gravity. It is screened within the K-mouflage radius
as $K' \rightarrow +\infty$ for $\chi \rightarrow -\infty$.

\item The propagation speed of scalar field fluctuations in the vacuum, around a spherically
symmetry object, is greater than $c$, and the static scalar field profile is linearly stable to
radial fluctuations.

\item Starting from a null initial condition, or a different profile, the scalar field relaxes to the
static profile. The evolution is regular for matter overdensities with a radius $R$ that is greater
than the K-mouflage radius $R_K$, whereas transient shocks appear when $R \lesssim R_K$.

\end{itemize}

\subsubsection{$K(\chi) \sim K_0 \chi^m$ with $K_0<0$ and $m$ odd, or $K_0>0$ and $m$ even, so that $K'<0$}

\begin{itemize}

\item The function $W_{-}(y)$ is not monotonically increasing up to $+\infty$ over $y>0$.
It has a maximum over $[0,+\infty[$ and goes to $-\infty$ for $y\rightarrow +\infty$.
There is no continuous static scalar field profile for high-density matter profiles.
One can build an infinite number of discontinuous static profiles, by patching together
disjoint intervals of $y$.

\item The hyperbolicity of the partial differential equations that govern the dynamics of the
scalar field is not guaranteed, and it generically breaks down in the nonlinear regime and for
high-density matter profiles. This means that in general the evolution with time of the system
is not well defined (this is no longer a Cauchy problem but an elliptic problem that requires
boundary conditions at late times). Thus, the discontinuous static solutions that can
be built are not physical.

\end{itemize}

Therefore, one must either disregard these models or include higher-order corrections to the
kinetic function $K(\chi)$.

\subsubsection{DBI$^{+}$ model}

\begin{itemize}

\item The function $W_{-}(y)$ is monotonically increasing up to $+\infty$ over $0\leq y < y_-$,
where $y_-$ is finite, and there is a unique
well-defined static scalar field profile for any static matter density profile.
The spatial gradients of the scalar field in any static state have a finite upper bound
set by $y_-$.

\item The fifth force amplifies Newtonian gravity. It is screened within the K-mouflage radius
as $K' \rightarrow +\infty$ for $\chi \rightarrow \chi_-$, with $\chi_- = - y_-^2/2$.

\item The propagation speed of scalar field fluctuations in the vacuum, around a spherically
symmetry object, is greater than $c$, and the static scalar field profile is linearly stable to
radial fluctuations.

\item Starting from a null initial condition, or a different profile, the scalar field relaxes to the
static profile. The evolution is regular for matter overdensities with a radius $R$ that is greater
than the K-mouflage radius $R_K$, whereas transient shocks appear when $R \lesssim R_K$.

\end{itemize}

\subsubsection{DBI$^{-}$ model}

\begin{itemize}

\item The function $W_{-}(y)$ is monotonically increasing up to a finite maximum $W_{\rm max}$
over $0\leq y < +\infty$.
For low-density matter fluctuations there is a unique
well-defined static scalar field profile, but for high-density matter profiles a static solution
can only be defined on separated inner and outer regions, $[0,r_-[$ and $]r_+,+\infty[$.

\item The fifth force amplifies Newtonian gravity. It is not screened within the K-mouflage radius
as $K' \rightarrow 0$ for $\chi \rightarrow -\infty$. In fact, the fifth force becomes much greater
than the Newtonian force in the nonlinear regime (until the model becomes ill-defined).

\item For moderate matter overdensities, where a static scalar field profile exists, the scalar
field relaxes to the latter. For high matter overdensities, where a static profile can only be
defined over $0 \leq r < r_-$ and $r>r_+$, the scalar field only relaxes to the static profile
on the outer region $r>r_+$. On smaller radii, the amplitude of the scalar field gradients
keeps increasing with time and no static solution can be reached.

\end{itemize}

Therefore, one must either disregard these models or include higher-order corrections to the
kinetic function $K(\chi)$.

\subsection{Healthy K-mouflage examples}

Finally, quantum mechanically, the theories with monomials in $\chi^m$ that appear with negative coefficients, viewed as low-energy effective theories,  cannot be embedded in a UV completion of the theory which satisfies the analyticity of the S-matrix \cite{Adams:2006aa}.
Hence only theories with odd powers of $\chi$  and $K_0>0$ are free of all these pathological behaviors.

More generally, avoiding ghosts requires $K'>0$ for $\chi>0$, while obtaining a realistic
small-scale behavior with efficient screening requires $K'>0$ for
$\chi<0$, with a sufficiently large value of $K'$ at large negative $\chi$.

To obtain a continuous and well-defined cosmological behavior up to high redshift,
the background Klein-Gordon equation, which reads in the matter-dominated era as
\beq
t> 0 : \;\; \dot{\bar\varphi} K'[ \dot{\bar\varphi}^2/(2{\cal M}^4) ] \simeq
- \frac{\beta \bar\rho t}{M_{\rm Pl}} ,
\label{KG-back}
\eeq
(where we made the approximation $\dd A/\dd\varphi \simeq \beta/M_{\rm Pl}$)
must admit a continuous solution. This requires that $W_{+}(y) \equiv y K'(y^2/2)$
increases monotonically to $+\infty$ over $y>0$, with $y=-\dot{\bar\varphi}/{\cal M}^2$
(where the subscript ``+'' in $W_+$ recalls the plus sign in the argument of $K'$).
In particular, this implies that $K'+2\chi K''>0$ for $\chi>0$.

To obtain a well-defined static scalar field profile for any matter density profile requires
that $W_{-}(y) = y K'(-y^2/2)$ is monotonically increasing to $+\infty$ over $y>0$.
In particular, this implies that $K'+2\chi K''>0$ for $\chi<0$.
This automatically means that the static profile is stable to radial fluctuations, which propagate
at a speed $c_s$ that is greater than $c$.
To ensure a well-defined evolution of the scalar field, for any configuration, we must have in
addition $K'+2\chi K''>0$ for all $\chi$.

Therefore, we find that healthy K-mouflage models should satisfy the conditions:
\beq
\chi_- < \chi < \chi_+ : \;\; K'(\chi) > 0 , \;\;\; K'+2\chi K''>0 ,
\label{cond-Kp-Ks}
\eeq
where $\chi_{\pm}=\pm y_{\pm}^2/2$ may be finite or $\pm\infty$, depending on the
domain of definition of $K(\chi)$, and over the range $0 \leq y < y_{\pm}$,
\beqa
W_{\pm}(y) = y K' \! \left( \pm \frac{y^2}{2} \right) &&
\mbox{are monotonically increasing} \nonumber \\
&& \mbox{to } +\infty .
\label{cond-Wp-Wm}
\eeqa
This implies in particular that $K(\chi)$ cannot be an even function of $\chi$,
whereas if $K(\chi)$ is odd, or more generally $K'(\chi)$ is even (which allows for an
additive constant that can play the role of the cosmological constant), the constraints arising
from the cosmological background and the small-scale static regime coincide.

A typical example that satisfies all these constraints is a cubic form such as $K=-1+\chi+\chi^3$.
As compared with the standard kinetic term $K=-1+\chi$ (with the cosmological constant
associated with the factor $-1$), the nonlinearities should
not distort too much the shape of the function $K(\chi)$ (i.e., avoid oscillations or
local maxima) but simply increase the derivative $K'$ for large $| \chi |$.

DBI-like kinetic functions, with $K(\chi) \supset \pm \sqrt{1\pm 2\chi}$, are either ill defined
in the cosmological domain, $\chi >0$ (they cannot follow a matter dominated cosmology
at early times), or in the small-scale static domain, $\chi <0$ (a static profile cannot be
defined for high matter overdensities).
A possible choice is to use a non-analytic kinetic function, such as
$K(\chi)= - \sqrt{1-2\chi}$ for $\chi>0$ and $K(\chi)= \sqrt{1+2\chi}-2$ for $\chi<0$.
Analytic functions with a similar behavior (i.e., $\chi$ is restricted to a finite range,
while $K'>0$, $K'+2\chi K''>0$, and $W_{\pm}$ are monotonically increasing to $+\infty$) are
for instance $K(\chi) = \arcsin(\chi)-1$, with $-1<\chi<1$, or $K(\chi) = \tan(\chi)-1$, with
$-\pi/2 < \chi < \pi/2$.

\section{Conclusion}
\label{sec:conclusion}

We have considered static configurations of K-mouflage models around compact objects. We have found that the dynamics of the Klein-Gordon equation
and the convergence to these solutions can only be reached  when the potential function
$W_{-}$ is monotonically increasing to infinity.
This fact is associated to the existence of real characteristic speeds for the Klein-Gordon equation and corresponds to travelling wave perturbations with a speed greater than the speed
of light for small perturbations around the static configurations. Such cases include cubic models with a bounded from below Lagrangian, and the wrong-sign DBI$^{+}$ models. When the potential
is bounded, and the compact object is screened, no convergence to a static configuration can be attained within the K-mouflage radius where spatial gradients of the scalar field diverge. This is what happens for the DBI$^{-}$ models.
Models with multiple extrema in their potential make no sense as the Klein-Gordon equation is not a well-defined hyperbolic equation as characteristic speeds become complex.

On the cosmological side, the potential function $W_{+}$ must also be monotonically increasing
to infinity. This implies that $K(\chi)$ cannot be even, whereas cosmological and
small-scale self-consistency conditions coincide when $K'(\chi)$ is even.

In addition to these theoretical self-consistency conditions, in order to obtain an efficient
screening mechanism for small high-density objects, the kinetic function must satisfy
$K'(\chi) \rightarrow +\infty$ for large negative $\chi$, or at least reach a very large value
so that the fifth force is suppressed by the factor $1/K'$ in the nonlinear regime.

We also note from Eq.(\ref{KG-omega-curl}) that in the general case (i.e., when the density
field is not spherically symmetric), the gradient $\nabla_{\vr}\varphi$ of the scalar field,
and the fifth force $\vF_{\varphi} \propto \nabla_{\vr}\varphi$, are not aligned with the
Newtonian force $\nabla_{\vr}\Psi_{\rm N}$, because the relationship between
$\nabla_{\vr}\varphi$ and $\nabla_{\vr}\Psi_{\rm N}$ involves an additional divergence-less
field ${\vec\omega}$ that arises from the rotational part of $\nabla_{\vr}\Psi_{\rm N}/K'$.

Our study shows that it is important to investigate the nonlinear small-scale regime of screened models of modified gravity in order
to guarantee that a theory is meaningful. The DBI$^{-}$ example is a prime example here.
Indeed it is well behaved on cosmological scales
but is meaningless, or at least incomplete, on small static scales.
In addition to the properties of the small scale static regime (e.g., checking that a static
scalar field profile exists for any matter overdensity), it is important to consider dynamical
properties, such as the relaxation of the scalar field. In particular, requiring that the dynamics
are well defined for any configuration (i.e., that we obtain a well-defined Cauchy problem) can
yield further constraints on the models.

Our results also show that in some cases models that would appear safe in a perturbative
approach are actually meaningless, and their flaws are not due to some approximation
scheme (e.g., making a quasistatic approximation) but to the nonlinearity of the model
that can give rise to complex behaviors (such as the absence of static states or
ill-defined Cauchy problems).

Here, we may note that numerical studies of other nonlinear models, involving nonlinear
derivative terms, such as the Galileon models, have faced problems as the numerical
algorithm encounters complex numbers during the evolution with time
\cite{Barreira:2013eea,Barreira:2013aa}.
Although this may be due to the quasistatic approximation used in these numerical schemes,
the analogy with the K-mouflage models suggests that the problem might be more serious
and signal a true shortcoming of these models, that could become ill-defined in some configurations.
We leave a detailed study of this point to future work.

\begin{acknowledgments}

This work is supported in part by the French Agence Nationale de la Recherche under Grant ANR-12-BS05-0002.

\end{acknowledgments}

\bibliography{ref1}   

\end{document}